\documentclass[preprint,12pt]{elsarticle}

\usepackage{amssymb,amsmath}\def\d{{\, \rm d}}

\usepackage{lineno,amsthm,comment}
\newtheorem{theorem}{Theorem}[section]
\newtheorem{proposition}[theorem]{Proposition}

\begin{document}

\begin{frontmatter}

\title{Lagrangian Descriptors with Uncertainty}

\author[label1]{Nan Chen\corref{cor1}}
\ead{chennan@math.wisc.edu}
\affiliation[label1]{organization={Department of Mathematics, University of Wisconsin-Madison},
            addressline={480 Lincoln Dr.},
            city={Madison},
            postcode={53706},
            state={WI},
            country={United States of America}}
            
\author[label2]{Evelyn Lunasin}
\affiliation[label2]{organization={Department of Mathematics, United States Naval Academy},
            addressline={Chauvenet Hall, 572C Holloway Road},
            city={Annapolis},
            postcode={21402-5002},
            state={MD},
            country={United States of America}}
            
\author[label2,label3]{Stephen Wiggins}
\affiliation[label3]{organization={School of Mathematics, University of Bristol},
            addressline={Fry Building, Woodland Road},
            city={Bristol},
            postcode={BS8 1UG},
            country={United Kingdom}}

\begin{abstract}
Lagrangian descriptors provide a global dynamical picture of the geometric structures for arbitrarily time-dependent flows with broad applications. This paper develops a mathematical framework for computing Lagrangian descriptors when uncertainty appears. The uncertainty originates from estimating the underlying flow field as a natural consequence of data assimilation or statistical forecast. It also appears in the resulting Lagrangian trajectories. The uncertainty in the flow field directly affects the path integration of the crucial nonlinear positive scalar function in computing the Lagrangian descriptor, making it fundamentally different from many other diagnostic methods. Despite being highly nonlinear and non-Gaussian, closed analytic formulae are developed to efficiently compute the expectation of such a scalar function due to the uncertain velocity field by exploiting suitable approximations. A rapid and accurate sampling algorithm is then built to assist the forecast of the probability density function (PDF) of the Lagrangian trajectories. Such a PDF provides the weight to combine the Lagrangian descriptors along different paths. Simple but illustrative examples are designed to show the distinguished behavior of using Lagrangian descriptors in revealing the flow field when uncertainty appears. Uncertainty can either completely erode the coherent structure or barely affect the underlying geometry of the flow field. The method is also applied for eddy identification, indicating that uncertainty has distinct impacts on detecting eddies at different time scales. Finally, when uncertainty is incorporated into the Lagrangian descriptor for inferring the source target, the likelihood criterion provides a very different conclusion from the deterministic methods.
\end{abstract}


\end{frontmatter}

\section{Introduction}

Complex nonlinear dynamical systems are ubiquitous in different scientific areas, including geophysics, climate science, engineering, neuroscience, and material science~\cite{wiggins1994normally, vallis2017atmospheric, strogatz2018nonlinear, wilcox1988multiscale, sheard2009principles, ghil2012topics}. They exhibit rich dynamical features, such as multiscale structures, intermittent instabilities, extreme events, and chaotic behavior \cite{farazmand2019extreme, trenberth2015attribution, moffatt2021extreme, majda2003introduction, manneville1979intermittency}. These complex nonlinear systems have been widely used to model and understand various natural phenomena. They also play an essential role in advancing many practical tasks, such as prediction, state estimation, and parameter inferences \cite{asch2016data, kalnay2003atmospheric, majda2012filtering, law2015data, ghil1991data}.

Lagrangian descriptor is a powerful tool for studying complex nonlinear dynamical systems. It provides a global dynamical picture of the geometric structures for arbitrarily time-dependent flows \cite{mendoza2010hidden}. The Lagrangian descriptor was initially developed in the context of geophysical fluid dynamics to analyze Lagrangian transport and mixing processes via identifying hyperbolic trajectories and their stable and unstable manifolds \cite{madrid2009distinguished, lopesino2017theoretical, mancho2013lagrangian}. It has several advantages over other trajectory diagnostics. First, it is computationally efficient and straightforward to implement. Second, it focuses on integrating a positive scalar function along trajectories of initial conditions of the system instead of tracking their phase space location. In this way, by emphasizing initial conditions, it directly targets the building blocks where the dynamical structure of phase space is encoded. The method has thus the capability of producing a complete and detailed geometrical phase space tomography in high dimensions by using lower dimensional phase space slices to extract the intersections of the phase space invariant manifolds with these slices \cite{demian2017detection, naik2019finding, naik2019finding, garcia2020tilting}. Third, all the invariant manifolds of the dynamical system are obtained simultaneously from the Lagrangian descriptor. In other words, the method can reveal all hyperbolic trajectories and their stable and unstable manifolds in a single calculation. In addition to facilitating computational efficiency, such a unique feature provides the input for the application of rigorous theorems such as the existence theorem for normally hyperbolic invariant manifolds (NHIMs) and their stable and unstable manifolds \cite{wiggins1994normally} and the Smale–Birkhoff homoclinic theorem for the existence of chaotic dynamics \cite{wiggins2003introduction}.

As a nonlinear dynamics tool to explore phase space, the Lagrangian descriptor has been widely applied in resolving many practical problems, including detecting  mesoscale eddies \cite{vortmeyer2016detecting}, assessing the predictive capacity of oceanic data sets \cite{mendoza2014lagrangian}, and tracing the origins of oil spill events in the Eastern Mediterranean \cite{garcia2022structured}. In addition to ocean science, the Lagrangian descriptor has essential applications in atmospheric studies, such as analyzing the structure of the Stratospheric Polar Vortex and its relation to sudden stratospheric warmings and ozone hole formation \cite{de2012routes, curbelo2019lagrangian, curbelo2019lagrangian2}. Recently, the Lagrangian descriptor has also received a significant amount of recognition in the field of chemistry to advance the computation of chemical reaction rates that are functions of the phase space structures \cite{craven2015lagrangian, craven2017lagrangian}. It facilitates the analysis of isomerization reactions \cite{garcia2020exploring, naik2020detecting} and the study of the influence of bifurcations on the manifolds that control chemical reactions \cite{garcia2020tilting}.

Due to the lack of a perfect understanding of nature, uncertainty exists in studying many complex nonlinear dynamical systems \cite{majda2012lessons, mignolet2008stochastic, majda2018model, majda2016introduction, kalnay2003atmospheric, palmer2001nonlinear, givon2004extracting, tremolet2007model}. In the presence of uncertainty, the states of the dynamical systems are no longer characterized deterministically. Instead, the probability density function (PDF) is utilized to describe each state variable. One primary source of the uncertainty comes from the inadequate characterization of small-scale features, which nevertheless impact the resolved state variables via nonlinear energy transfer. Specifically, when random noises or stochastic parameterizations are adopted to describe the statistical behavior of these unresolved-scale variables, the uncertainty arises naturally in the time evolution of the model trajectories \cite{palmer2001nonlinear, majda2012lessons, orrell2001model, benner2015survey}. Another source of uncertainty comes from  inaccurate measurements in many practical situations when data is utilized to assist the estimation of model states \cite{evensen2009data, law2015data}. It consists of  observational noise when the state variables are directly observed. It also includes the inference errors in recovering the unobserved variables from noisy observations when only indirect, sparse, or coarse-grained measurements are available. The uncertainties resulting from the observational measures and the underlying dynamical systems significantly affect the state estimation, which consequently influences the calculation of Lagrangian trajectories.

Obtaining the exact spatiotemporal velocity field is often a prerequisite in computing the Lagrangian descriptor to reveal the underlying flow structures. In studying many ocean science problems, the velocity field can be inferred accurately from the observed sea surface height (SSH) \cite{qiu2020reconstructing, liu2005patterns, doglioni2021sea}. However, sea-surface heights are usually converted to velocities by using the assumption of geostrophic balance. The ageostrophic component brings about uncertainties in the inferred flow field. In addition, satellite observations of the SSH are not always available in certain ocean regions. For example, in the marginal ice zone (MIZ) of the Arctic area, the presence of sea ice floes prevent a direct inference of the ocean field \cite{manucharyan2017submesoscale, covington2022bridging}. In such a situation, Lagrangian data assimilation \cite{apte2013impact, apte2008data, apte2008bayesian, ide2002lagrangian, chen2014information}, which exploits the observed ice floes, becomes essential for recovering the ocean velocity field. Uncertainty arrives as a natural consequence of such an estimated ocean field in the form of the so-called posterior distribution. Note that purely data-driven approaches, such as clustering and sequential Monte Carlo methods, have been developed to estimate the large-scale ocean structures exploiting merely the trajectory data \cite{maclean2017coherent, hadjighasem2016spectral, froyland2015rough}. Yet, if a dynamical or surrogate model is used to provide additional information on the underlying turbulent flow field and assists with these trajectory data through Lagrangian data assimilation, then the inference of the coherent structure is expected to be improved. In general, uncertainty is inevitable when the underlying turbulent flow field is not perfectly known. Data assimilation becomes essential to reduce, but not entirely eliminate, the uncertainty in the resulting estimated flow field by combining limited observations with suitable approximate models in a wide range of geophysical and engineering problems \cite{kalnay2003atmospheric, lahoz2010data, majda2012filtering, evensen2009data, law2015data}. Uncertainty may also appear in the initialization of the underlying flow field, and it can be significantly amplified in the subsequent forecast of turbulent signals. In the presence of uncertainty, the resulting statistical forecast becomes very different from the deterministic one based on a single trajectory. Therefore, incorporating these uncertainties into calculating the Lagrangian descriptor is essential to characterize the additional features beyond the deterministic trajectories.

This paper develops a general mathematical framework for computing Lagrangian descriptors when uncertainty appears. The uncertainty originates from estimating the underlying flow field as a natural consequence of data assimilation or statistical forecast. Since the Lagrangian trajectories are driven by the flow velocity, uncertainty also appears in forecasting these trajectories. Both types of uncertainty affect the computation of the Lagrangian descriptor, and they are handled in different ways. On the one hand, the path integration of the crucial nonlinear positive scalar function in computing the Lagrangian descriptor depends on the flow velocity field, making it fundamentally different from many other diagnostic methods. Despite being highly nonlinear and non-Gaussian, closed analytic formulae are developed to efficiently compute the expectation of such a scalar function due to the uncertain velocity field by exploiting suitable approximations. On the other hand, a rapid and accurate sampling method is developed to extract the time evolution of the velocity fields from the posterior distribution from the Lagrangian data assimilation. The sampled velocity fields are adopted to assist the forecast of the possible range of the Lagrangian trajectories. The resulting PDF of the locations where the Lagrangian trajectories lie provides the weight to combine the Lagrangian descriptors along different paths that leads to the final value of the Lagrangian descriptor. As closed analytic formulae are designed to analyze both sources of uncertainty, the new Lagrangian descriptor framework is computationally efficient and stable. It is worth mentioning that applying uncertainty quantification to analyzing flow structures has been discussed in several works. A recent study consists of a systematic comparison between nine different methods in detecting the Lagrangian coherent structures when uncertainties appear in trajectory data \cite{badza2023sensitive}. Many other studies focused on the uncertainty quantification of the finite-time Lyapunov exponent (FTLE) type of methods \cite{schneider2011variance, guo2016finite, bozorgmagham2015atmospheric, you2021computing, balasuriya2020uncertainty, rapp2020uncertain}. Some recent work also discussed the uncertainties in using the Lagrangian descriptors \cite{garcia2022bridge, garcia2023new}. Nevertheless, unlike most existing approaches emphasizing the uncertainty in the Lagrangian trajectories, the method developed here has a unique feature of characterizing the uncertainties in both the velocity fields and the Lagrangian trajectories from a probabilistic viewpoint. The former is extremely important as it directly affects the crucial nonlinear positive scalar function in computing the Lagrangian descriptor. The method developed here is also naturally adaptive to the solution from data assimilation, allowing to combine noisy data with dynamical or statistical models for uncertainty quantification. The framework will be utilized to study the new features appearing in the identified geometric structures of the flow characterized by the Lagrangian descriptor when uncertainty arises. Simple, but illustrative, examples will be exploited to understand these new geometric features related to several scientific topics, including identifying eddies at different time scales using the solution from data assimilation, and detecting the source of a given target. The former is crucial for understanding the transport and mixing of the fluid and the impact on marine biology, while the latter has important implications in facilitating the understanding of many environmental issues such as the oil spill problem.

The rest of the paper is organized as follows. A mathematical modeling framework of the random flow field is described in Section \ref{Sec:Flow_Model}. It also includes an efficient Lagrangian data assimilation method and the statistical forecast with uncertainty quantification. Section \ref{Sec:LD} consists of developing the Lagrangian descriptor in the presence of uncertainty and the associated computational methods. Section \ref{Sec:Numerics} includes a set of examples that reveals the inferred geometric features resulting from the Lagrangian descriptor when the uncertainty is taken into consideration.

\section{Modeling Random Flow Field, Lagrangian Data Assimilation, and Uncertainty Quantification}\label{Sec:Flow_Model}
Characterizing the underlying flow field is the prerequisite for applying the Lagrangian descriptor. This section provides a mathematical framework for modeling random flow fields with uncertainty quantification. The general framework developed here is computationally efficient and mathematically tractable. It can describe many complex dynamical systems in practice and serve as a general testbed.
\subsection{A mathematical framework of modeling random flow field}
The randomness in the underlying flow field is one of the main sources that induce the uncertainty in the Lagrangian descriptor. The randomness comes from the insufficient resolution of small-scale features and the intrinsic behavior of turbulent flows. To develop a general framework for modeling random flow field that is both mathematically tractable and practically applicable, consider the following spectral representation of the underlying flow velocity field \cite{chen2015noisy, majda2003introduction},
\begin{equation}\label{Ocean_Velocity}
  \mathbf{u}(\mathbf{x},t) = \sum_{\mathbf{k}\in\mathcal{K},\alpha\in\mathcal{A}}\hat{u}_{\mathbf{k},\alpha}(t)e^{i\mathbf{k}\mathbf{x}}\mathbf{r}_{\mathbf{k},\alpha},
\end{equation}
where $\mathbf{x}=(x,y)^\mathtt{T}$ is the two-dimensional coordinate. A double periodic domain is adopted here, and the flow field is given by a finite summation of Fourier modes. The index $\mathbf{k}=(k_1,k_2)^\mathtt{T}$ is the wavenumber, and the index $\alpha$ represents different types of waves, including, for example, the gravity modes and the geophysically balanced modes in the study of many geophysical flows. The set $\mathcal{K}$ usually consists of all the wavenumbers that satisfy $-K_{\mbox{max}}\leq k_1, k_2\leq K_{\mbox{max}}$ with $K_{\mbox{max}}$ being an integer that is pre-determined. The vector $\mathbf{r}_{\mathbf{k},\alpha}$ is the eigenvector, which links the two components of velocity fields, namely $u$ and $v$. For conciseness of notations, the explicit dependence of $\alpha$ of $\hat{u}_{\mathbf{k},\alpha}$ and $\mathbf{r}_{\mathbf{k},\alpha}$ in \eqref{Ocean_Velocity} is omitted in the following discussions. In other words, the Fourier coefficient and the eigenvector are simply written as $\hat{u}_{\mathbf{k}}$ and $\mathbf{r}_{\mathbf{k}}$. Since the left hand side of \eqref{Ocean_Velocity} is evaluated at physical space, the Fourier coefficients $\hat{u}_{\mathbf{k}}$ and $\hat{u}_{-\mathbf{k}}$ for all $\mathbf{k}$ are complex conjugates. So do the eigenvectors $\mathbf{r}_{\mathbf{k}}$ and $\mathbf{r}_{-\mathbf{k}}$. Note that the Fourier basis functions are adopted here to simplify the description of the framework. Different basis functions and boundary conditions can be utilized in \eqref{Ocean_Velocity} for various applications in practice. Therefore, the representation in \eqref{Ocean_Velocity} is general.

Stochastic models are used to describe the time evolution of each Fourier coefficient $\hat{u}_{\mathbf{k}}$ in \eqref{Ocean_Velocity}, which is a much more computationally efficient way to mimic the observed turbulent flows generated from a complicated PDE system. The stochastic model is often calibrated by matching several key statistics in the observed time series of $\hat{u}_{\mathbf{k}}$. Among different stochastic models, the linear stochastic model, namely the complex Ornstein-Uhlenbeck (OU) process \cite{gardiner1985handbook}, is a widely used choice:
\begin{equation}\label{OU_process}
  \frac{\d\hat{u}_{\mathbf{k}}}{\d t} = (- d_\mathbf{k} + i\omega_\mathbf{k}) \hat{u}_{\mathbf{k}} + \mathbf{f}(t) + \sigma_\mathbf{k}\dot{W}_\mathbf{k},
\end{equation}
where $d_\mathbf{k}, \omega_\mathbf{k}$ and $\mathbf{f}(t)$ are damping, phase and deterministic forcing, $\sigma_\mathbf{k}$ is the noise coefficient and $\dot{W}_\mathbf{k}$ is a white noise. The constants $d_\mathbf{k}$, $\omega_\mathbf{k}$ and $\sigma_\mathbf{k}$ are real-valued while the forcings are complex. The stochastic noise in the linear stochastic model is utilized to effectively parameterize the nonlinear deterministic time evolution of chaotic or turbulent dynamics \cite{majda2016introduction, farrell1993stochastic, berner2017stochastic, branicki2018accuracy, majda2018model, li2020predictability, harlim2008filtering, kang2012filtering}.

The mathematical framework of modeling the random flow field with linear stochastic models characterizing the time series of spectral modes in \eqref{Ocean_Velocity}--\eqref{OU_process} has been widely used to describe various turbulent flow fields, including the rotating shallow water equation \cite{chen2015noisy} and the quasi-geostrophic equation \cite{chen2023stochastic}. It has also been adopted as an effective surrogate forecast model in data assimilation to approximate the Navier-Stokes equations \cite{branicki2018accuracy}, moisture-coupled tropical waves \cite{harlim2013test} and a nonlinear topographic barotropic model \cite{chen2023uncertainty}. Quantifying the uncertainty using the linear stochastic model as a surrogate model in the statistical forecast and filtering can be found in \cite{branicki2013non, chen2023uncertainty, chen2016model}.

Starting from a Gaussian initial condition, the statistics of $\hat{u}_{\mathbf{k}}$ in \eqref{OU_process} remain Gaussian. This facilitates uncertainty quantification, which requires the information of only the leading two moments: mean and variance.

\subsection{Uncertainty in the statistical forecast}
Forecasting the flow field is a prerequisite for many practical situations. Due to the turbulent nature, the statistical forecast becomes essential for obtaining future states, where the forecast PDF provides a natural way for quantifying the uncertainty. For the complex OU process \eqref{OU_process}, the time evolutions of the mean $\overline{\hat{u}_{\mathbf{k}}}$ and the variance $\mbox{var}(\hat{u}_{\mathbf{k}})$ of $\hat{u}_{\mathbf{k}}$ can be written by closed analytic formulae \cite{majda2012filtering}:
\begin{equation}\label{OU_mean_var}
\begin{aligned}
  \overline{\hat{u}_{\mathbf{k}}}(t) &= \overline{\hat{u}_{\mathbf{k}}}(t_0)e^{(- d_\mathbf{k} + i\omega_\mathbf{k})(t-t_0)} + \int_{t_0}^tf(s)e^{(- d_\mathbf{k} + i\omega_\mathbf{k})(t-s)}\d s,\\
  \mbox{var}(\hat{u}_{\mathbf{k}})(t) &= \mbox{var}(\hat{u}_{\mathbf{k}})(t_0) e^{-2d_\mathbf{k}(t-t_0)} + \frac{\sigma_\mathbf{k}^2}{2d_\mathbf{k}}\left(1-e^{2d_\mathbf{k}(t-t_0)}\right),
\end{aligned}
\end{equation}
where $t_0$ is the initial time and $t$ is the time for the statistical forecast. Initial mean $\overline{\hat{u}_{\mathbf{k}}}(t_0)$ and initial variance $\mbox{var}(\hat{u}_{\mathbf{k}})(t_0)$ are needed for such a statistical forecast. The initial uncertainty is zero if the initial value is perfectly known. The initial distribution may also come from data assimilation, which will be discussed in the following subsection.

\subsection{Uncertainty in state estimation via data assimilation}\label{Subsec:LD}
\subsubsection{Overview of data assimilation}
In practice, the state of the velocity field is often obtained by combining observational data with a numerical model. Without any information from observations, the most reasonable estimation of the state is given by the equilibrium distribution of the model. However, the equilibrium state usually contains considerable uncertainty due to the intrinsic chaotic or turbulent features. Data assimilation, which optimally combines model output with available noisy observational data, is a widely used method to reduce uncertainty and provides an improved state estimation. The fundamental principle of data assimilation is the Bayes theorem. The model output is known as the prior distribution, while the observations give the likelihood. Their product leads to the so-called posterior distribution that serves as the solution for data assimilation. The general ideas and rigorous mathematical derivations for data assimilation can be found in literature, for example, \cite{asch2016data, kalnay2003atmospheric, majda2012filtering, law2015data, ghil1991data}.

\begin{figure}
\centering
  \hspace*{-1cm}\includegraphics[width=15cm]{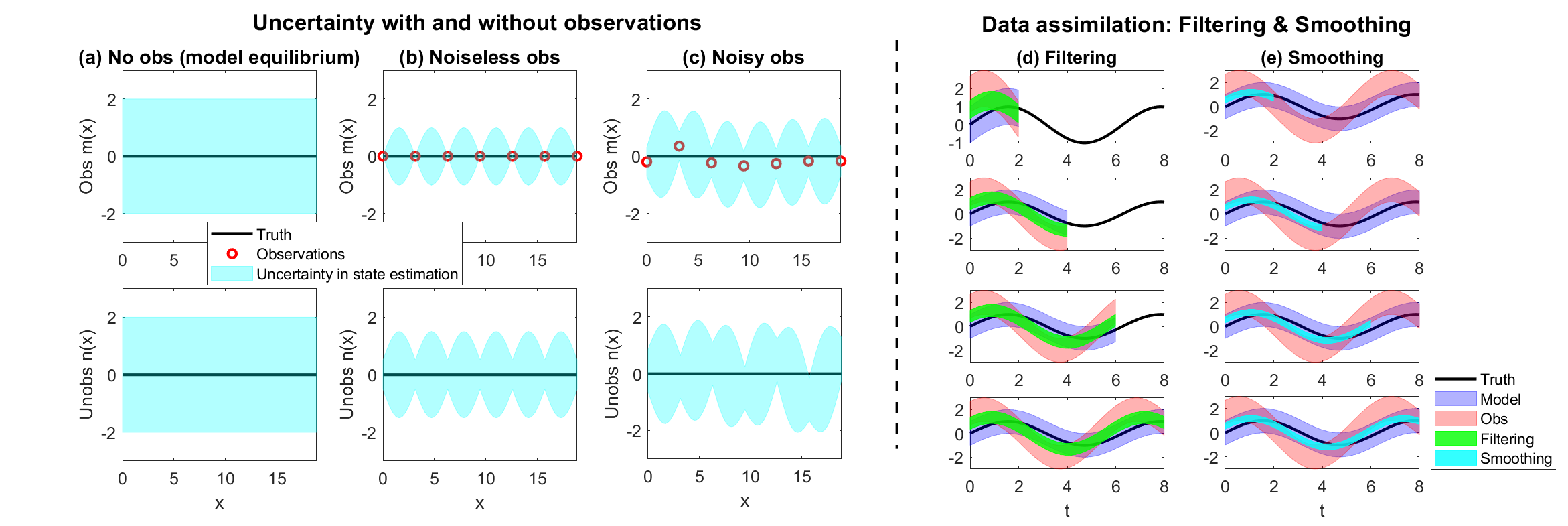}
  \caption{Panels (a)--(c): Schematic illustration of the basic idea of data assimilation that combines model output with data in state estimation. Panels (d)--(e): Schematic illustration of filtering and smoothing, which are two data assimilation approaches.   }\label{fig: DA_Illustration}
\end{figure}

Panels (a)--(c) of Figure \ref{fig: DA_Illustration} include a schematic illustration of state estimation via data assimilation. Consider two state variables $m(x)$ and $n(x)$, where $x$ is the coordinate. These variables are coupled via an underlying turbulent dynamical system. Panel (a) shows their equilibrium distribution, which contains uncertainty (cyan area). Now assume observations (red dots) are available at a few discrete points. Further assume the observations are only available for the variable $m$ but not for $n$. If the observations are perfect, as is shown in Panel (b), then the state estimation of $m(x)$ at the observational locations is estimated perfectly. The state estimation of $m(x)$ at the locations between these observations still contains uncertainty, but the uncertainty is smaller than that at the model equilibrium due to the spatial correlation of the system such that the observations have an impact on the entire $m(x)$. Similarly, the correlation between the observed and unobserved variables reduces the uncertainty in $n(x)$ with the help of the available discrete observations in $m(x)$. Therefore, observations play a role as constraints to the model output that provides additional information from the model and reduces the uncertainty in state estimation. In practice, the observations are typically polluted by noise or representation error \cite{janjic2018representation}. Therefore, the state estimations are imperfect even at the observational locations. See Panel (c). Bayes theorem thus gives the optimal state estimation solution that accounts for the uncertainties in both model and observations.

In practice, data assimilation is carried out sequentially in time as the underlying system is often given by a dynamical model. Filtering and smoothing are two different data assimilation approaches. See Panel (d)--(e) of Figure \ref{fig: DA_Illustration}. Filtering only exploits observational information in the past to estimate the current state. The filtering solution, also known as the filtering posterior distribution, is thus utilized as the initialization for the subsequent real-time forecast. In contrast, smoothing is more widely used as a postprocessing method. After obtaining the observational information within an interval $[0,T]$, state estimation is carried out at each time instant $t$ within the interval $[0,T]$, which is also a typical technique for obtaining reanalysis data in climate science \cite{uppala2005era, kalnay2003atmospheric}. As additional ``future'' information beyond time $t$ is used in smoothing, its state estimation is expected to be more accurate than filtering in the sense that the mean estimation contains less error and the uncertainty shrinks as well. Filtering is running forward, while smoothing contains a forward (filtering) run from $0$ to $T$ and then a backward run from $T$ to $t$ to further reduce the uncertainty. Filtering can be regarded as a prerequisite for smoothing. In the remainder of this paper, data assimilation always means smoothing, as the goal is to analyze the uncertainty within a given interval instead of a real-time forecast. The framework can be easily applied to the filtering solution for relevant applications.

\subsubsection{A mathematical tractable framework for Lagrangian data assimilation}
When the velocity field is directly observed at fixed grid points, they are known as Eulerian observations. Assume the observational operator is linear and the observational noise is Gaussian. In such a case, if the stochastic models in \eqref{OU_mean_var} for different $\mathbf{k}$ are used as the forecast model in data assimilation, then the solution of $\hat{u}_{\mathbf{k}}$ is given by a Gaussian distribution from the Kalman filter \cite{kalman1960new}. If a set of nonlinear forecast models is utilized, then the solution provided by the ensemble data assimilation schemes can also be approximately by Gaussian distributions. On the other hand, Lagrangian data assimilation is widely used for recovering the ocean flow field. Lagrangian data assimilation exploits drifters to provide trajectory data that can be used to infer the underlying velocity field \cite{apte2013impact, apte2008data, apte2008bayesian, ide2002lagrangian}. The observational process of Lagrangian data assimilation is highly nonlinear. Nevertheless, with the linear stochastic models in \eqref{OU_mean_var} being the surrogate forecast model, analytic solutions are available for the state estimation, significantly facilitating the computation \cite{chen2014information}. Denote the observational process by
\begin{equation}\label{Tracer_eqn}
\frac{\d\mathbf{x}}{\d t} = \mathbf{u}(\mathbf{x},t) + \sigma_\mathbf{x}\mathbf{W}_\mathbf{x},
\end{equation}
where $\mathbf{W}_\mathbf{x}$ is a two-dimensional real-valued white noise representing the observational uncertainty and small-scale perturbations to the observed tracers while $\sigma_\mathbf{x}$ is the noise coefficient. The velocity field $\mathbf{u}$ in \eqref{Tracer_eqn} is given by \eqref{Ocean_Velocity}, which is a highly nonlinear function of $\mathbf{x}$. Usually, there are $L$ equations of \eqref{Tracer_eqn} that are used in Lagrangian data assimilation, representing the observed trajectories of $L$ Lagrangian tracers.

Define $\mathbf{X}=(\mathbf{x}_1,\ldots,\mathbf{x}_L)^\mathtt{T}$ the collection of the $L$ observed tracer trajectories and $\mathbf{U}=\{\hat{u}_\mathbf{k}\}$ the vector that collects the Fourier coefficients. In light of \eqref{Ocean_Velocity}, \eqref{OU_process} and \eqref{Tracer_eqn}, the Lagrangian data assimilation can be written in the following form.
\begin{proposition}[Posterior distribution of Lagrangian data assimilation: Filtering]
Given one realization of the tracer trajectories $\mathbf{X}(s\leq t)$, the filtering posterior distribution $p(\mathbf{U}(t)|\mathbf{X}(s\leq t))$ of Lagrangian data assimilation \eqref{eq:cgns} is conditionally Gaussian,
 where the time evolutions of the conditional mean $\boldsymbol\mu$ and the conditional covariance $\bf R$ are given by
\begin{subequations}\label{eq:cgns}
\begin{align}
\frac{\d \mathbf{X}(t)}{\d t} &= \mathbf{A}(\mathbf{X}, t) \mathbf{U}(t) + \sigma_\mathbf{x} \dot{\mathbf{W}}_\mathbf{X}(t),\label{eq:cgns_X}\\
\frac{\d \mathbf{U}(t)}{\d t} &=  \mathbf{F}_\mathbf{U} + \boldsymbol{\Lambda} \mathbf{U}(t)  + \boldsymbol{\Sigma}_\mathbf{U} \dot{\mathbf{W}}_\mathbf{U}(t),\label{eq:cgns_U}
\end{align}
\end{subequations}
where $\mathbf{A}(\mathbf{X}, t)$ contains all the Fourier bases and is, therefore, a highly nonlinear function of $\mathbf{X}$.
\begin{subequations}\label{eq:filter}
\begin{align}
\frac{\d\boldsymbol{\mu}}{\d t} &= \left(\mathbf{F}_\mathbf{U} + \boldsymbol{\Lambda} \boldsymbol{\mu}\right)  + \sigma_\mathbf{x}^{-2}\mathbf{R}\mathbf{A}^\ast\left(\frac{\d \mathbf{X}}{\d t} - \mathbf{A}\boldsymbol{\mu} \right),\label{eq:filter_mu}\\
\frac{\d\mathbf{R}}{\d t} &= \boldsymbol{\Lambda}\mathbf{R} + \mathbf{R}\boldsymbol{\Lambda}^\ast + \boldsymbol{\Sigma}_\mathbf{U}\boldsymbol{\Sigma}_\mathbf{U}^\ast - \sigma_x^{-2}\mathbf{R}\mathbf{A}^\ast\mathbf{A}\mathbf{R},\label{eq:filter_R}
\end{align}
\end{subequations}
with $\cdot^*$ being the complex conjugate transpose.
\end{proposition}
With the filtering solution \eqref{eq:filter} in hand, closed analytic formulae are also available for the smoothing solution.
\begin{proof}
The proof can be found in \cite{liptser2013statistics, chen2018conditional}.
\end{proof}

\begin{proposition}[Posterior distribution of Lagrangian data assimilation: Smoothing]\label{Prop:Smoothing}
Given one realization of the tracer trajectories $\mathbf{X}(t)$ for $t\in[0,T]$, the smoother estimate $p(\mathbf{U}(t)|\mathbf{X}(s), s\in[0,T])\sim\mathcal{N}(\boldsymbol\mu_\mathbf{s}(t),\mathbf{R}_\mathbf{s}(t))$ of the coupled system is also Gaussian,
where the conditional mean $\boldsymbol\mu_\mathbf{s}(t)$ and conditional covariance $\mathbf{R}_\mathbf{s}(t)$ of the smoother satisfy the following backward equations
\begin{subequations}\label{Smoother_Main}
\begin{align}
  \frac{\overleftarrow{\d \boldsymbol{\mu}_\mathbf{s}}}{\d t} &=  -\mathbf{F}_\mathbf{U} - \boldsymbol\Lambda\boldsymbol{\mu}_\mathbf{s}  + (\boldsymbol{\Sigma}_\mathbf{U}\boldsymbol{\Sigma}_\mathbf{U}^*)\mathbf{R}^{-1}(\boldsymbol\mu - \boldsymbol{\mu}_\mathbf{s}),\label{Smoother_Main_mu}\\
  \frac{\overleftarrow{\d \mathbf{R}_\mathbf{s}}}{\d t} &= - (\boldsymbol\Lambda + (\boldsymbol{\Sigma}_\mathbf{U}\boldsymbol{\Sigma}_\mathbf{U}^*) \mathbf{R}^{-1})\mathbf{R}_\mathbf{s} - \mathbf{R}_\mathbf{s}(\boldsymbol\Lambda^* + (\boldsymbol{\Sigma}_\mathbf{U}\boldsymbol{\Sigma}_\mathbf{U}^*)\mathbf{R})  + \boldsymbol{\Sigma}_\mathbf{U}\boldsymbol{\Sigma}_\mathbf{U}^* ,\label{Smoother_Main_R}
\end{align}
\end{subequations}
with $\boldsymbol\mu$ and $\mathbf{R}$ being given by \eqref{eq:filter}. The notation $\overleftarrow{\d \cdot}/\d t$ corresponds to the negative of the usual derivative, which means that the system \eqref{Smoother_Main} is solved backward over $[0,T]$ with the starting value of the nonlinear smoother being the same as the filter estimate $(\boldsymbol\mu_\mathbf{s}(T), \mathbf{R}_\mathbf{s}(T)) = (\boldsymbol\mu(T), \mathbf{R}(T))$.
\end{proposition}

The smoother estimate \eqref{Smoother_Main} provides a PDF at each time instant for the recovered velocity field, which includes the uncertainty. Given these PDFs and the temporal dependence, an efficient sampling algorithm of the time series of the velocity field $\mathbf{U}$ from the posterior distributions can be developed. The sampled time series of the velocity field will be used to forecast the possible range of the Lagrangian trajectories $\mathbf{x}(t)$ in computing the Lagrangian descriptor.
\begin{proof}
The proof can be found in
 \cite{chen2020learning}.
\end{proof}

\begin{proposition}[Sampling trajectories from posterior distributions]\label{Prop:Sampling}
Based on the smoother estimate, an optimal backward sampling of the trajectories associated with the unobserved variable $\mathbf{U}$ satisfies the following explicit formula,
\begin{equation}\label{Sampling_Main}
  \frac{\overleftarrow{\d \mathbf{U}}}{\d t} = \frac{\overleftarrow{\d \boldsymbol\mu_\mathbf{s}}}{\d t} - \big(\boldsymbol\Lambda + (\boldsymbol{\Sigma}_\mathbf{U}\boldsymbol{\Sigma}_\mathbf{U}^*)\mathbf{R}^{-1}\big)(\mathbf{U} - \boldsymbol\mu_\mathbf{s}) + \boldsymbol{\Sigma}_\mathbf{U}\dot{\mathbf{W}}_{\mathbf{U}}(t).
\end{equation}
\end{proposition}
\begin{proof}
The proof can be found in
 \cite{chen2020learning}.
\end{proof}

The temporal dependence in the sampled time series of $\mathbf{U}$ is extremely important. It contains the memory effect of the recovered velocity field, which is a crucial dynamical feature that affects the prediction of the Lagrangian trajectories $\mathbf{x}(t)$. The sampling approach in \eqref{Sampling_Main} fundamentally differs from drawing independent samples at different time instants, which essentially gives a noisy time series that lacks the physical properties of $\mathbf{U}$.
\subsection{Converting the uncertainty from spectral space to grid points in physical space}
Recall in \eqref{Ocean_Velocity} that the velocity field is represented in the spectral form. Denote by $\overline{\hat{u}}_\mathbf{k}$ the mean and $\mbox{var}(\hat{u}_\mathbf{k})$ the variance of mode $\mathbf{k}$, where the mean and variance can be those from the forecast \eqref{OU_mean_var} or from the data assimilation \eqref{Smoother_Main}. Note that in the result from the Lagrangian data assimilation, the entire posterior covariance $\mathbf{R}_{\mathbf{s}}$ is, in general, not a non-diagonal matrix due to the mixing of the modes in the observation process. Nevertheless, the diagonal components of $\mathbf{R}_{\mathbf{s}}$ usually have more significant amplitudes than the off-diagonal ones, especially when the estimation of $\mathbf{U}$ becomes more accurate \cite{chen2014information}. Therefore, taking the diagonal entries, which represent the actual uncertainty of each mode, to reconstruct the variance at a grid point in physical space is a natural and reasonable choice. The following argument utilizes the mean-fluctuation decomposition of each Gaussian random variable $\hat{u}_\mathbf{k} = \overline{\hat{u}}_\mathbf{k} + \hat{u}_\mathbf{k}^\prime$, where $\overline{\hat{u}}_\mathbf{k}$ is the mean and $\hat{u}_\mathbf{k}^\prime$ is the fluctuation with $\mbox{var}(\hat{u}_\mathbf{k}^\prime) = \mbox{var}(\hat{u}_\mathbf{k})$.

The mean at each grid point is given by
\begin{equation}\label{Ocean_Velocity_mean}
  \overline{u}(\mathbf{x},t) = \sum_{\mathbf{k}\in\mathcal{K}}\overline{\hat{u}}_{\mathbf{k}}(t)e^{i\mathbf{k}\mathbf{x}}\mathbf{r}_{\mathbf{k},1} \qquad \overline{v}(\mathbf{x},t) = \sum_{\mathbf{k}\in\mathcal{K}}\overline{\hat{u}}_{\mathbf{k}}(t)e^{i\mathbf{k}\mathbf{x}}\mathbf{r}_{\mathbf{k},2},
\end{equation}
where $\mathbf{r}_{\mathbf{k},1}$ and $\mathbf{r}_{\mathbf{k},2}$ are the two component of the eigenvector $\mathbf{r}_{\mathbf{k}}$. Similarly, the fluctuation in physical space is given by
\begin{equation}\label{Ocean_Velocity_fluctuation}
  {u}^\prime(\mathbf{x},t) = \sum_{\mathbf{k}\in\mathcal{K}}{\hat{u}}^\prime_{\mathbf{k}}(t)e^{i\mathbf{k}\mathbf{x}}\mathbf{r}_{\mathbf{k},1} \qquad {v}^\prime(\mathbf{x},t) = \sum_{\mathbf{k}\in\mathcal{K}}{\hat{u}}^\prime_{\mathbf{k}}(t)e^{i\mathbf{k}\mathbf{x}}\mathbf{r}_{\mathbf{k},2}.
\end{equation}
Due to the negligible off-diagonal components in the covariance matrix, the variance at a fixed location $\mathbf{x}$ and time $t$ is given by
\begin{equation}\label{Ocean_Velocity_fluctuation}
  \mbox{var}({u}(\mathbf{x},t)) = \sum_{\mathbf{k}\in\mathcal{K}}\mbox{var}(\hat{u}_{\mathbf{k}}(t))(\mathbf{r}_{\mathbf{k},1}\mathbf{r}^*_{\mathbf{k},1}) \qquad \mbox{var}({v}(\mathbf{x},t)) = \sum_{\mathbf{k}\in\mathcal{K}}\mbox{var}(\hat{u}_{\mathbf{k}}(t))(\mathbf{r}_{\mathbf{k},2}\mathbf{r}^*_{\mathbf{k},2}).
\end{equation}
Note that the variances at different grid points are the same. This is a unique feature when global basis functions such as the Fourier bases are used. When ensemble data assimilation methods are applied, the so-called localization technique for recovering the state variables in physical space becomes essential to eliminate the spurious correlations due to the sampling error \cite{petrie2008localization, houtekamer2005ensemble}. In such a case, the spatial distribution of the variance is usually not uniform. Such an inhomogeneous variance distribution is entirely due to numerical approximations. The data assimilation framework developed here avoids such an issue, and the resulting variance represents the exact uncertainty from the Bayesian inference. It is worthwhile to foreshadow that even though the variances at different grid points are the same when the Lagrangian data assimilation described in Section \ref{Subsec:LD} is adopted, the change in the Lagrangian descriptor as a result of such uncertainties is inhomogeneous in space. The change in the Lagrangian descriptor relies on the dynamical properties, and the uncertainty affects the Lagrangian descriptor in a highly nonlinear way.

\section{Lagrangian Descriptor in the Presence of Uncertainty}\label{Sec:LD}
\subsection{The standard Lagrangian descriptor with deterministic flow field}
Denote by $\mathbf{x}=(x,y)^\mathtt{T}$ the two-dimensional displacement and $\mathbf{u}=(u,v)^\mathtt{T}$ the two-dimensional velocity field.
The general formula of the Lagrangian descriptor is as follows \cite{mancho2013lagrangian, lopesino2017theoretical, garcia2022lagrangian}
\begin{equation}\label{LD_General_Formula}
  \mathcal{L}(\mathbf{x}^*,t^*) = \int_{t-\tau}^{t+\tau} F(\mathbf{x}, t) \d t,
\end{equation}
where $F=|\tilde{F}|$ is a scalar field with positive values and $t$ is time. According to \eqref{LD_General_Formula}, $\mathcal{L}$ is the integrated modulus of $\tilde{F}$
along a trajectory from the past $t-\tau$ to the future $t+\tau$ that goes through a point $\mathbf{x}^*$ at time $t^*$. This way yields a space- and time-dependent field computed for all $\mathbf{x}^*$ and $t^*$. One commonly used Lagrangian descriptor is by taking $F$ to be the arc length of the path traced by the trajectory. That is,
\begin{equation}\label{LD_VelocityBased}
M_{vel}(\mathbf{x}^*,t^*) = \int_{t-\tau}^{t+\tau} \sqrt{\left(\frac{\partial x}{\partial t}\right)^2+\left(\frac{\partial y}{\partial t}\right)^2} \d t = \int_{t-\tau}^{t+\tau} \sqrt{u^2+v^2} \d t.
\end{equation}
Once the Lagrangian descriptor is computed, it is usually normalized to its maximum value in space for illustration purposes. Several other Lagrangian descriptors have also been widely used in practice. One is analog to \eqref{LD_VelocityBased} but exploits the vorticity instead of the arc length in defining $F$. Such a vorticity-based Lagrangian descriptor is essential for identifying vortex-like patterns, such as the eddy detection in the ocean. Another Lagrangian descriptor takes the direct difference between $F$ at the current and a former time instant. It is a useful metric to identify the source of a given target, which is crucial for tracing the source of the oil split and many other environmental problems. Incorporating the uncertainty into these Lagrangian descriptors will be discussed in Sections \ref{Subsec:VorticityLD} and \ref{Subsec:SourceDetectionLD}.

\begin{figure}[htb]
\centering
  \hspace*{-2cm}\includegraphics[width=10cm]{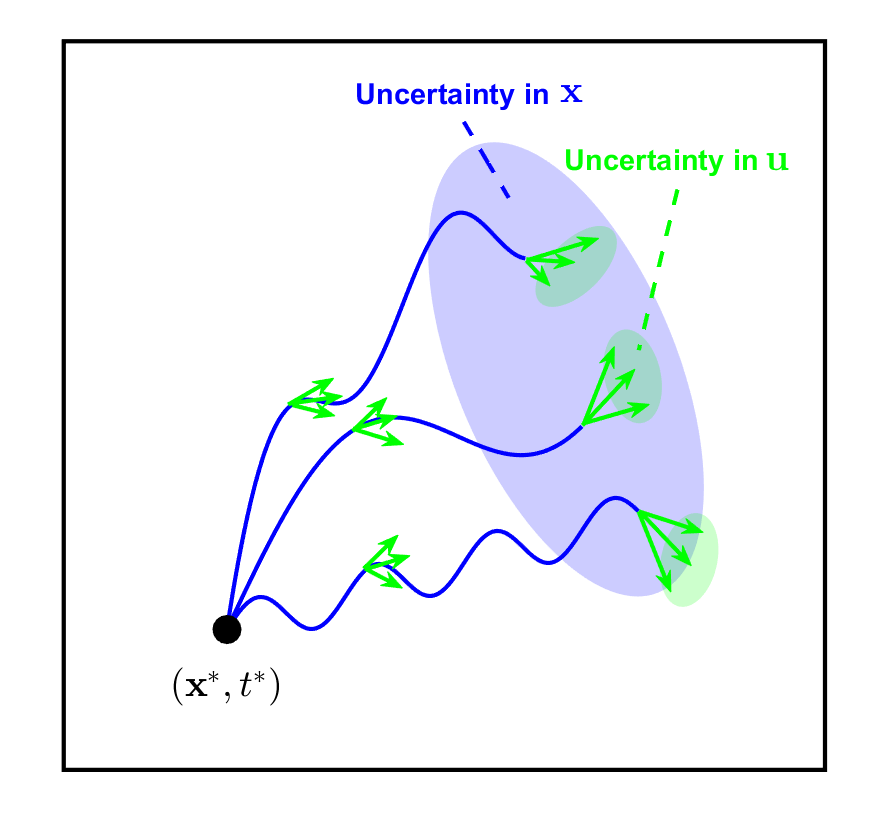}
  \caption{A schematic illustration of the two sources of uncertainties affecting the calculation of the Lagrangian descriptor: the uncertainty in the estimated velocity field $\mathbf{u}$ (green shading area) and the resulting uncertainty in forecasting the Lagrangian trajectory $\mathbf{x}$ (blue shading area). }\label{fig: Uncertainty_Illustration}
\end{figure}

The expression in \eqref{LD_VelocityBased} explicitly depends on the velocity field while the integration is alone the Lagrangian trajectory. In the standard definition in \eqref{LD_General_Formula}, the velocity field $\mathbf{u}$ and the trajectory $\mathbf{x}$ are both deterministic. This happens when $\mathbf{u}$ is accurately inferred from observations. Yet, in many practical situations, the velocity field $\mathbf{u}$ coming from reanalysis or estimations contains uncertainty. Consequently, the displacement $\mathbf{x}$, driven by $\mathbf{u}$, also becomes non-deterministic. These are the two primary sources of uncertainty in computing the Lagrangian descriptor. See Figure \ref{fig: Uncertainty_Illustration} for an illustration of these two sources of uncertainties.

\subsection{Lagrangian descriptor with uncertainty}
The uncertainty usually comes directly from estimating the velocity field $\mathbf{u}$. Notably, since the governing equation of the trajectory is given by Newton's law
\begin{equation}\label{Governing_equation_x}
\frac{\d\mathbf{x}}{\d t} = \mathbf{u}(\mathbf{x},t),
\end{equation}
the uncertainty appears in $\mathbf{x}$ as well. Sometimes, small noise can be added to \eqref{Governing_equation_x} \cite{badza2023sensitive, rapp2020uncertain, chen2014information}. It accounts for additional uncertainties due to contributions from small or unresolved scales. It should not be confused with the uncertainty of the velocity field due to measurement or inference that directly affects $\mathbf{x}$ in \eqref{Governing_equation_x} and is usually the dominant part of the uncertainty. If $\mathbf{u}$ at each time $t$ is given by a distribution that accounts for the uncertainty, then the forward and backward paths of $\mathbf{x}$ using \eqref{Governing_equation_x} also contain uncertainty and are characterized by distributions. It is worth highlighting that, despite being related, these two sources of uncertainties play different roles in calculating the Lagrangian descriptor. The uncertainty in the velocity field directly impacts computing the path integration of the positive scalar function $F=\sqrt{u^2+v^2}$. In contrast, the uncertainty in the trajectory leads to a PDF that provides the weight to combine the above positive quantity along different paths.

Therefore, in the presence of uncertainty, the Lagrangian descriptor in \eqref{LD_VelocityBased} should be modified by taking the expectation in terms of both $\mathbf{u}$ and $\mathbf{x}$. The expectation of $\mathbf{x}$ accounts for the uncertainty of where the Lagrangian trajectories are located. The expectation of $\mathbf{u}$ is for evaluating the integrand $\sqrt{u(\mathbf{x},t)^2+v(\mathbf{x},t)^2}$ in computing Lagrangian descriptor at each possible fixed location. See Figure \ref{fig: Uncertainty_Illustration} again. The former is natural, which has been considered in the previous works for analyzing Lagrangian coherent structures using different Eulerian and Lagrangian methods \cite{badza2023sensitive, rapp2020uncertain}. The latter is unique for the Lagrangian descriptor, which has not been incorporated into such a trajectory diagnostic approach in the existing work. Taking into account these uncertainties, the Lagrangian descriptor is given by
\begin{equation}\label{LD_VelocityBased_UQ}
\begin{aligned}
M_{vel}^{UQ}(\mathbf{x}^*,t^*) &= \mathbb{E}_{\mathbf{u},\mathbf{x}}\left[\int_{t-\tau}^{t+\tau} \sqrt{u(\mathbf{x},t)^2+v(\mathbf{x},t)^2} \d t\right]\\
&= \int_{t-\tau}^{t+\tau} \int_{\mathbf{x}}\int_{\mathbf{u}}\sqrt{u(\mathbf{x},t)^2+v(\mathbf{x},t)^2}p(\mathbf{u}|\mathbf{x})p(\mathbf{x}) \d\mathbf{u}\d\mathbf{x}\d t.
\end{aligned}
\end{equation}
Note that $\mathbf{u}$ is a function of $\mathbf{x}$ as the velocity depends on the location. Once $t$ is given, the distribution of $\mathbf{u}$ is obtained from state estimation (i.e., from statistical forecast or data assimilation). On the other hand, computing the probability of the forward or backward path $\mathbf{x}$ depends on the initial condition $\mathbf{x}^*$, time $t$ and the underlying velocity field $\mathbf{u}$. Thus, \eqref{LD_VelocityBased_UQ} is rewritten as
\begin{equation}\label{LD_VelocityBased_UQ_2}
M_{vel}^{UQ}(\mathbf{x}^*,t^*) = \int_{t-\tau}^{t+\tau} \left(\int_{\mathbf{x}}\mathbb{E}_{\mathbf{u}}\left[\sqrt{u(\mathbf{x},t)^2+v(\mathbf{x},t)^2}\right]p(\mathbf{x}) \d\mathbf{x}\right)\d t.
\end{equation}
A direct calculation of the Lagrangian descriptor in \eqref{LD_VelocityBased_UQ_2} via Monte Carlo simulation is computationally quite expensive as the integration in \eqref{LD_VelocityBased_UQ_2} is at least in a 5-dimensional space $(x,y,u,v,t)$, where $u$ and $v$ are further given by a much higher dimensional system. In addition, the Lagrangian descriptor needs to be computed at each grid point of the initial value $\mathbf{x}^*$ in the two-dimensional physical space. Therefore, the focus below is on developing an approximate solution with analytic formulae to efficiently compute the Lagrangian descriptor in \eqref{LD_VelocityBased_UQ_2}.

\subsubsection{Coping with the uncertainty coming from the underlying flow}
The goal here is to compute
\begin{equation}\label{LD_VelocityBased_UQ_2_partI}
  \mathcal{F}_E(\mathbf{x},t) = \mathbb{E}_{\mathbf{u}}\left[\sqrt{u(\mathbf{x},t)^2+v(\mathbf{x},t)^2}\right]
\end{equation}
in \eqref{LD_VelocityBased_UQ_2}.
Now assume the velocity field $u(\mathbf{x},t)$ and $v(\mathbf{x},t)$ contain uncertainties at each fixed location $\mathbf{x}$ and fixed time instant $t$. For simplicity, assume both $u$ and $v$ at fixed location and time are Gaussian distributed such that
\begin{equation}\label{mean_fluctuation_decomposition}
u = \overline{u} + u^\prime \qquad\mbox{and}\qquad v = \overline{v} + v^\prime,
\end{equation}
where $\overline{u}$ and $\overline{v}$ are two numbers, representing the posterior mean, and $u^\prime\sim\mathcal{N}(0,\sigma_u^2)$ and $v^\prime\sim\mathcal{N}(0,\sigma_v^2)$ are two zero-mean Gaussian random variables, representing the posterior variances in the context of data assimilation or forecast described in Section \ref{Sec:Flow_Model}. For notation simplicity, the dependence of $u$ and $v$ on $\mathbf{x}$ and $t$ are not explicitly written, but both velocity components take values at a fixed time and location. This assumption is reasonable as many data assimilation methods provide approximate Gaussian posterior distributions. Although there is no closed analytic solution of $\mathcal{F}_E(\mathbf{x},t)$, the following proposition leads to an approximate solution that can efficiently compute $\mathcal{F}_E(\mathbf{x},t)$.

\begin{proposition}\label{Proposition_1}
In light of the mean-fluctuation decomposition in \eqref{mean_fluctuation_decomposition}, define $B:=(u')^2+(v')^2$. Then, $\mathcal{F}_E(\mathbf{x},t)$ in \eqref{LD_VelocityBased_UQ_2_partI} is given by
\begin{equation}\label{Formula_F_E}
  \mathcal{F}_E(\mathbf{x},t) \approx \sqrt{\left(\sqrt{\bar{u}^2 + \bar{v}^2} + \mathbb{E}\sqrt{B}\right)^2 - 2\sqrt{\bar{u}^2 + \bar{v}^2}\cdot\mathbb{E}\sqrt{B}},
\end{equation}
where
\begin{equation}\label{Approx_B}
  \mathbb{E}\sqrt{B}\approx \frac{\Gamma \left(\frac{(\sigma_u+\sigma_v)^2}{2(\sigma_u^2+\sigma_v^2) + \frac{1}{2}}\right)}{\Gamma\left(\frac{(\sigma_u+\sigma_v)^2}{2(\sigma_u^2+\sigma_v^2)}\right)}\sqrt{2\frac{\sigma_u^2+\sigma_v^2}{\sigma_u+\sigma_v}}
\end{equation}
with $\Gamma(\cdot)$ being the Gamma function.
\end{proposition}
The derivation of the results in Proposition \ref{Proposition_1} is included in Appendix. The analytic formulae in Proposition \ref{Proposition_1} facilitate an efficient approach to cope with the uncertainty resulting from the velocity field.

\begin{figure}[htb]\centering
  \hspace*{-1cm}\includegraphics[width=16cm]{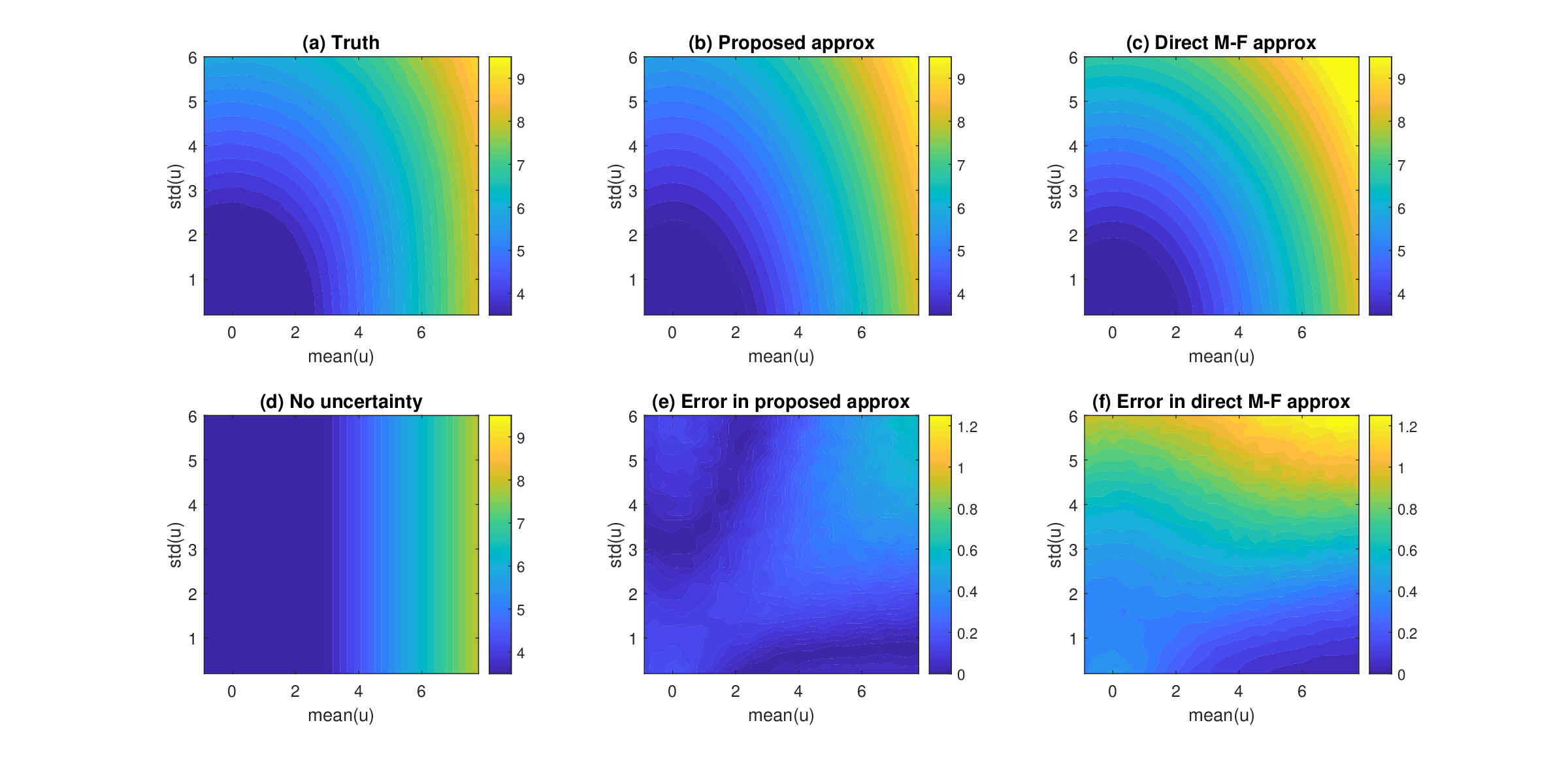}
  \caption{Numerical validation of Proposition \ref{Proposition_1}. Panels (a)--(d): computing the $\mathcal{F}_E$ using different methods. Panel (a): The true solution based on a Monte Carlo simulation with $N_{MC} = 50,000$ samples. Panel (b): The proposed approximation of computing $\mathcal{F}_E(\mathbf{x},t)$ using the formulae \eqref{Formula_F_E}--\eqref{Approx_B} in Proposition \ref{Proposition_1}. Panel (c): A direct mean-fluctuation (M-F) approximation shown in \eqref{mean_fluc_approximation_F_E} by interchanging the expectation and the square root. Panel (d): The approximation by completely dropping the uncertainty in \eqref{no_uncertainty_F_E}. Note that the color bars for these four panels are the same. Panels (e)-(f): the absolute error between the truth and the two methods shown in Panels (b)--(c), respectively, where the color bars in these two panels are also set to be the same. }\label{fig: Gamma_approx_validation}
\end{figure}

Numerical validation of Proposition \ref{Proposition_1} is provided here. For simplicity, the validation test is carried out at a fixed time and location. Therefore, $u$ and $v$ are both one-dimensional random variables. Let $v$ be a Gaussian variable with mean $\overline{v}=2$ and standard deviation $\mbox{std}(v) = 2$. Figure \ref{fig: Gamma_approx_validation} shows the result of $\mathcal{F}_E=\mathbb{E}_{\mathbf{u}}\left[\sqrt{u^2+v^2}\right]$ as a function of the mean and the standard deviation of $u$. Panels (a)--(d) show the computed $\mathcal{F}_E$ using four different methods. Panel (a) shows the truth, computed based on a Monte Carlo simulation with $N_{MC} = 50,000$ samples. That is, one sample of $u$ and one sample of $v$ are drawn from the above Gaussian distributions in each run to compute $\sqrt{u^2+v^2}$, and then the average is taken over all the runs. Panel (b) shows the proposed approximation of computing $\mathcal{F}_E(\mathbf{x},t)$ using the formulae \eqref{Formula_F_E}--\eqref{Approx_B} in Proposition \ref{Proposition_1}. Panel (c) includes the results using the formula:
\begin{equation}\label{mean_fluc_approximation_F_E}
\begin{aligned}
  \mathcal{F}_E &= \mathbb{E}_{\mathbf{u}}\left[\sqrt{u^2+v^2}\right]\approx
  \sqrt{\mathbb{E}_{\mathbf{u}}\left[u^2+v^2\right]}\\
  &=\sqrt{\overline{u}^2+\overline{v}^2+\overline{u'u'}^2+\overline{v'v'}^2}
  = \sqrt{\overline{u}^2+\overline{v}^2+\mbox{var}(u)+\mbox{var}(v)},
\end{aligned}
\end{equation}
which is a direct mean-fluctuation (M-F) approximation. Note that this direct method interchanges the expectation and the square root. Therefore, certain errors are expected, despite the simplicity of the formula. Panel (d) shows the approximation by completely dropping the uncertainty, namely,
\begin{equation}\label{no_uncertainty_F_E}
  \mathcal{F}_E = \mathbb{E}_{\mathbf{u}}\left[\sqrt{u^2+v^2}\right]\approx
  \sqrt{\overline{u}^2+\overline{v}^2},
\end{equation}
which is expected to have a significant error if the uncertainty is large. Panels (e)-(f) include the absolute error between the truth and the two methods shown in Panels (b)--(c), respectively. The comparison here indicates the following conclusions. First, considering the uncertainty makes a significant difference in computing $\mathcal{F}_E$ and thus the resulting Lagrangian descriptor. Second, the method in Proposition \ref{Proposition_1} outperforms a direct mean-fluctuation approximation by interchanging the expectation and the square root, especially when the uncertainty is large. Third, the error in the approximation methods increases monotonically as the uncertainty increases. Nevertheless, the method developed in Proposition \ref{Proposition_1} remains to have a small error even in the presence of large uncertainty. The results confirm that the proposed efficient method in Proposition \ref{Proposition_1} is a suitable approximation.

\subsubsection{Coping with the uncertainty in the Lagrangian trajectory}\label{Subsec:UQ_Trajectory}

With the $F_E(\mathbf{x},t)$ being computed, what remains is to estimate $p(\mathbf{x})$ to finish calculating the Lagrangian descriptor \eqref{LD_VelocityBased_UQ_2}. Different from the velocity field, the distribution of the trajectory at a given time instant $p(\mathbf{x})$ is generally non-Gaussian. This can be seen by noting that the domain has finite support while the support of a Gaussian distribution is infinite. In addition, the governing equation \eqref{Governing_equation_x} is not linear, which does not guarantee that the resulting distribution is Gaussian. Therefore, unlike the way to handle the underlying velocity field, applying a direct mean-fluctuation decomposition assuming a Gaussian distribution for $p(\mathbf{x})$ is inappropriate. In general, $p(\mathbf{x})$ can be estimated by first sampling $\mathbf{u}$ and then plugging the resulting $\mathbf{u}$ to \eqref{Governing_equation_x} that gives a set of $\mathbf{x}$. A two-dimensional kernel density estimation is then used to provide an analytic expression of $p(\mathbf{x})$. In Section \ref{Sec:Numerics}, a standard kernel density estimation with Gaussian kernels is adopted. The bandwidth is given by the rule-of-thumb bandwidth estimator for the two dimensions independently \cite{silverman1986density}.

In the context of the Lagrangian data assimilation in Section \ref{Subsec:LD}, the sampled time series of $\mathbf{u}$ can be obtained using the analytic formula in Proposition \ref{Prop:Sampling}. If the focus is on the statistical forecast, then a direct integration of \eqref{OU_process} can create a set of sampled time series of $\mathbf{u}$. Since $\mathbf{u}$ is written in spectral form, sampling or forecasting the coefficients for different modes can be carried out independently.

\subsection{The vorticity-based Lagrangian descriptor with uncertainty}\label{Subsec:VorticityLD}
The vorticity-based Lagrangian descriptor is a practically useful approach to characterize vortex-like behavior, such as eddies and other mesoscale features in the ocean \cite{vortmeyer2016detecting}, which has been shown to have several advantages over the eddy detection methods using Eulerian measurements \cite{vortmeyer2019comparing}.
Vorticity is the curl of the velocity field:
\begin{equation}
\omega(\mathbf{x},t) = \nabla\times \mathbf{u}(\mathbf{x},t) = \frac{\partial v(\mathbf{x},t)}{\partial x} - \frac{\partial u(\mathbf{x},t)}{\partial y}.
\end{equation}
A natural way to define a vorticity-based Lagrangian descriptor is the following:
\begin{equation}\label{LD_VorticityBased}
M_{vor}(\mathbf{x}^*,t^*) = \int_{t-\tau}^{t+\tau} \sqrt{\omega^2(\mathbf{x},t)} \d t = \int_{t-\tau}^{t+\tau} |\omega(\mathbf{x},t)| \d t,
\end{equation}
As an analog to \eqref{LD_VelocityBased_UQ}, when the vorticity is estimated with uncertainty, the corresponding LD for \eqref{LD_VorticityBased} is given by
\begin{equation}\label{LD_VorticityBased_UQ}
M^{UQ}_{vor}(\mathbf{x}^*,t^*) = \mathbb{E}_{\omega,\mathbf{x}}\left[\int_{t-\tau}^{t+\tau} |\omega(\mathbf{x},t)| \d t\right],
\end{equation}
The goal here is to cope with uncertainty in the integrand associated with the vorticity (which is related to the velocity field):
\begin{equation}\label{LD_VorticityBased_UQ_2_partI}
  \mathcal{F}_E(\mathbf{x},t) = \mathbb{E}_{\omega}\left[|\omega(\mathbf{x},t)|\right]
\end{equation}
while computing the uncertainty in the trajectory remains the same as those in Section \ref{Subsec:UQ_Trajectory}. The following proposition can be exploited to calculate $\mathcal{F}_E(\mathbf{x},t)$ in \eqref{LD_VorticityBased_UQ_2_partI}.
\begin{proposition}
Assume $\omega(\mathbf{x},t)\sim\mathcal{N}(\mu_\omega, \sigma_\omega^2)$ satisfies a Gaussian distribution, then $|\omega(\mathbf{x},t)|$ satisfies a folded Gaussian distribution and its expectation is
\begin{equation}\label{Expectation_Folded_Gaussian}
  \mathcal{F}_E(\mathbf{x},t) = \sigma_\omega\sqrt{\frac{2}{\pi}}e^{-\mu_\omega^2/(2\sigma_\omega^2)} + \mu_\omega \left(1-2\Phi\left(-\frac{\mu_\omega}{\sigma_\omega}\right)\right),
\end{equation}
where $\Phi$ is the normal cumulative distribution function $\Phi(x) = \frac{1}{2}\left[1+\mbox{erf}\left(\frac{x}{\sqrt{2}}\right)\right]$ with $\mbox{erf}$ being the error function \cite{tsagris2014folded}.
\end{proposition}

\subsection{Lagrangian descriptor for detecting the target source with uncertainty}\label{Subsec:SourceDetectionLD}
\begin{figure}[htb]\centering
  \hspace*{-1cm}\includegraphics[width=16cm]{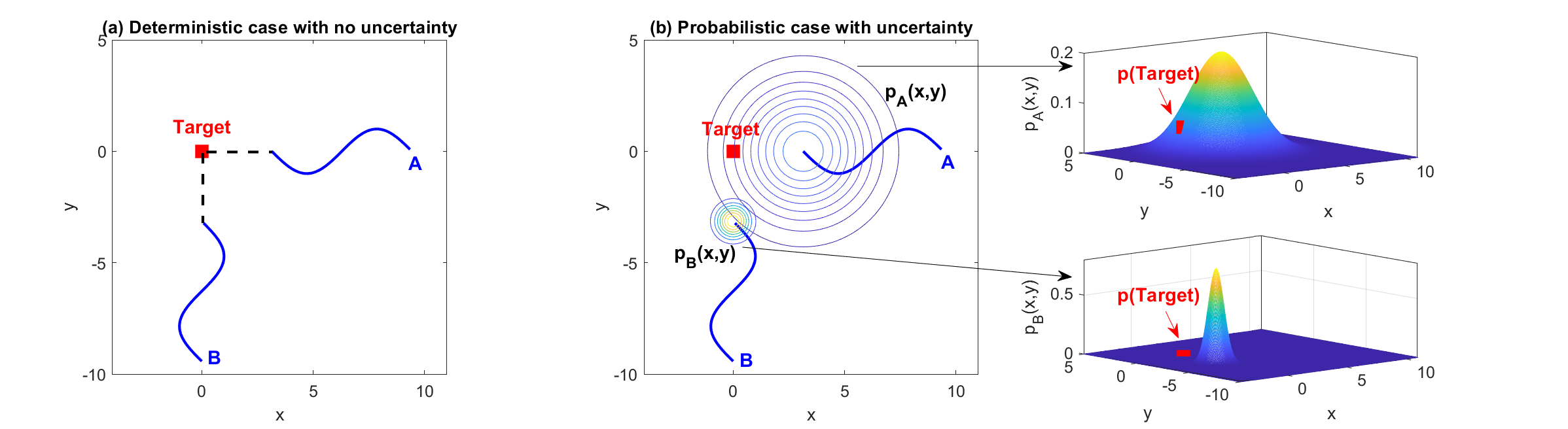}
  \caption{Schematic illustration of the Lagrangian descriptors for detecting the target source when the flow field is deterministic \eqref{LD_Source_Deterministic} (Panel (a)) and when the flow field contains uncertainty such that the forecast of the Lagrangian trajectories is characterized by a distribution \eqref{LD_Source_Probability} (Panel (b)). The distances from the forecast locations originating from points A and B to the target are the same in Panel (a). The likelihoods of the target using the two distributions in Panel (b) are, however, different.}\label{fig: Source_target_illustration}
\end{figure}

In \cite{garcia2023new}, a slightly different Lagrangian descriptor was introduced to detect the source of a given target. The method has broad applications. For example, it was used to study oil spills and many other environmental problems \cite{garcia2022structured}. The definition of such a Lagrangian descriptor is essentially given by the distance between the source and the target. It can be defined as follows:
\begin{equation}\label{LD_Source_Deterministic}
  L_B(\mathbf{x}, t^*, \tau) = \|\mathbf{x}(t^*-\tau)-\mathbf{x}^*\|,
\end{equation}
which aims to identify a target source $\mathbf{x}^*$ that is consistent with a later observation $\mathbf{x}$ at time $t$. The target source $\mathbf{x}^*$ is located at an earlier time $t^*-\tau$. The source is given by the $\mathbf{x}$ that minimizes $L_B(\mathbf{x}, t^*, \tau)$. When the underlying flow field is deterministic, each $\mathbf{x}$ gives a value of the corresponding $L_B(\mathbf{x}, t^*, \tau)$ in \eqref{LD_Source_Deterministic}. Note that different norms can be used to define $L_B(\mathbf{x}, t^*, \tau)$ \cite{garcia2022structured}, but the essence is to compute a certain path-wise distance between the predicted target and the actual target. See the dashed black lines in Panel (a) of Figure \ref{fig: Source_target_illustration}.

Now consider the situation that uncertainty is incorporated in the underlying flow field. Starting from $t^*-\tau$, the forecast location $\mathbf{x}$ at $t$ is now given by a distribution $p(\mathbf{x}(t))$. Therefore, the deterministic Lagrangian descriptor \eqref{LD_Source_Deterministic} is modified by considering the probability of the target $\mathbf{x}^*$ under $p(\mathbf{x}(t))$, where $p(\mathbf{x}(t))$ is the forecast PDF from a point value $\mathbf{x}(t^*-\tau)$,
\begin{equation}\label{LD_Source_Probability}
  L_B^{UQ}(\mathbf{x}, t^*, \tau) = p(\mathbf{x}^*).
\end{equation}
By starting from different locations, the probability gives a manifold that can be used to determine the most likely source of the target. See the illustration in Panel (b) of Figure \ref{fig: Source_target_illustration}. Note in this illustration that the path-wise distances from the predicted targets to the actual ones are the same, starting from A and B. However, the two forecast distributions are different. Therefore, the actual target has different probabilities evaluated by these two distributions. This indicates that using a probabilistic way to define the Lagrangian descriptor can lead to a very different conclusion from the deterministic approaches.

\section{Applications of Lagrangian descriptors with uncertainty}\label{Sec:Numerics}
\subsection{Two simple dynamical systems: flow fields with a quadrupole structure and a dipole structure}
Let us start with understanding the additional features resulting from the Lagrangian descriptor in the presence of uncertainty. To this end, two simple steady incompressible flow fields are considered here.

\begin{figure}[htb]\centering
  \hspace*{-1cm}\includegraphics[width=16cm]{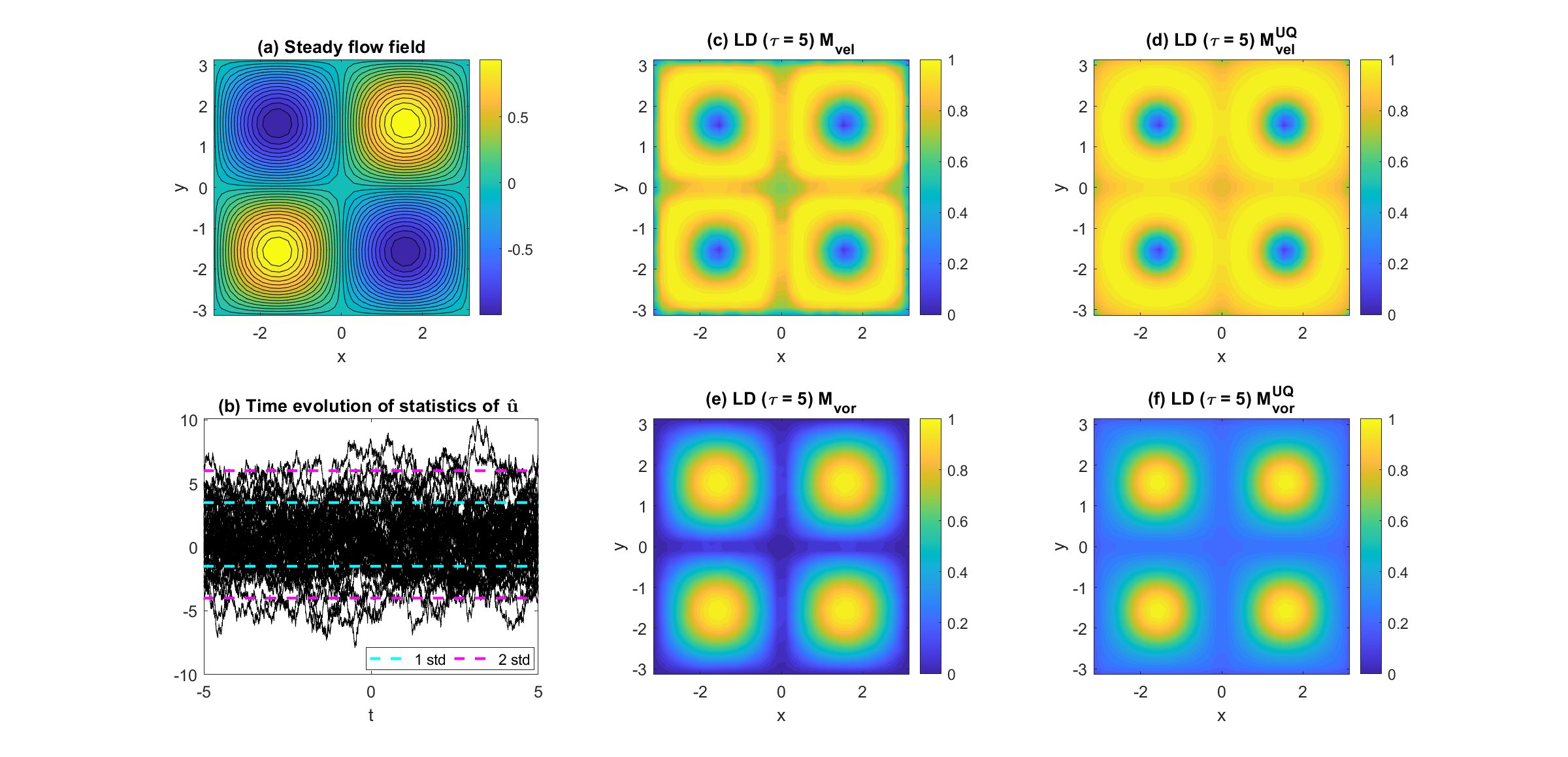}
  \caption{The Lagrangian descriptors for the flow field with the quadrupole structure. Panel (a): The true steady flow field using stream lines satisfied by \eqref{Quadrupole_flow} with $\hat{u}\equiv1$. Panel (b): The statistical equilibrium state of $\hat{u}$ (dashed lines) and $N_{MC}=50$ sampled trajectories (solid black curves). Panel (c): Velocity-based Lagrangian descriptor (LD) $M_{vel}$ using \eqref{LD_VelocityBased}. Panel (d): Vorticity-based Lagrangian descriptor $M_{vor}$ using \eqref{LD_VorticityBased}. Panel (e): Velocity-based Lagrangian descriptor (LD) with uncertainty $M^{UQ}_{vel}$ using \eqref{LD_VelocityBased_UQ}. Panel (f): Vorticity-based Lagrangian descriptor (LD) with uncertainty $M^{UQ}_{vor}$ using \eqref{LD_VorticityBased_UQ}. In computing all the Lagrangian descriptors, the value in determining the bounds of the integration is chosen as $\tau=5$.  }\label{fig: quadrupole_LD}
\end{figure}

The first flow field is given by a single mode, where the stream function $\psi$ and the two velocity components $(u, v)$ are given by
\begin{equation}\label{Quadrupole_flow}
\begin{aligned}
  \psi(x,y,t) &= -\hat{u}(t)\sin(x)\sin(y),\\
   u(x,y,t) &= \hat{u}(t)\sin(x)\cos(y),\qquad\mbox{and}\\
   v(x,y,t) &= -\hat{u}(t)\cos(x)\sin(y),
\end{aligned}
\end{equation}
respectively. For the true flow field, the coefficient $\hat{u}$ is assumed to be a constant $\hat{u}(t)\equiv 1$. This leads to a quadrupole structure in the flow field. See Panel (a) of Figure \ref{fig: quadrupole_LD}. This is a typical example used to study various methods in identifying Lagrangian coherent structures \cite{schlueter2019model, vortmeyer2016detecting}. Uncertainty is introduced by assuming $\hat{u}$ is not perfectly known. Instead, it is driven by a real-valued OU process:
\begin{equation}\label{Quadrupole_SDE}
  \frac{\d\hat{u}}{\d t} = -d_u \hat{u} + f_u + \sigma_u\dot{W}_u.
\end{equation}
Here the parameters are given by
\begin{equation}\label{Quadrupole_parameters}
  d_u = 0.5,\qquad\sigma_u = 2.5\qquad \mbox{and} \qquad f_u = 0.5.
\end{equation}

\begin{figure}[htb]\centering
  \hspace*{-1cm}\includegraphics[width=16cm]{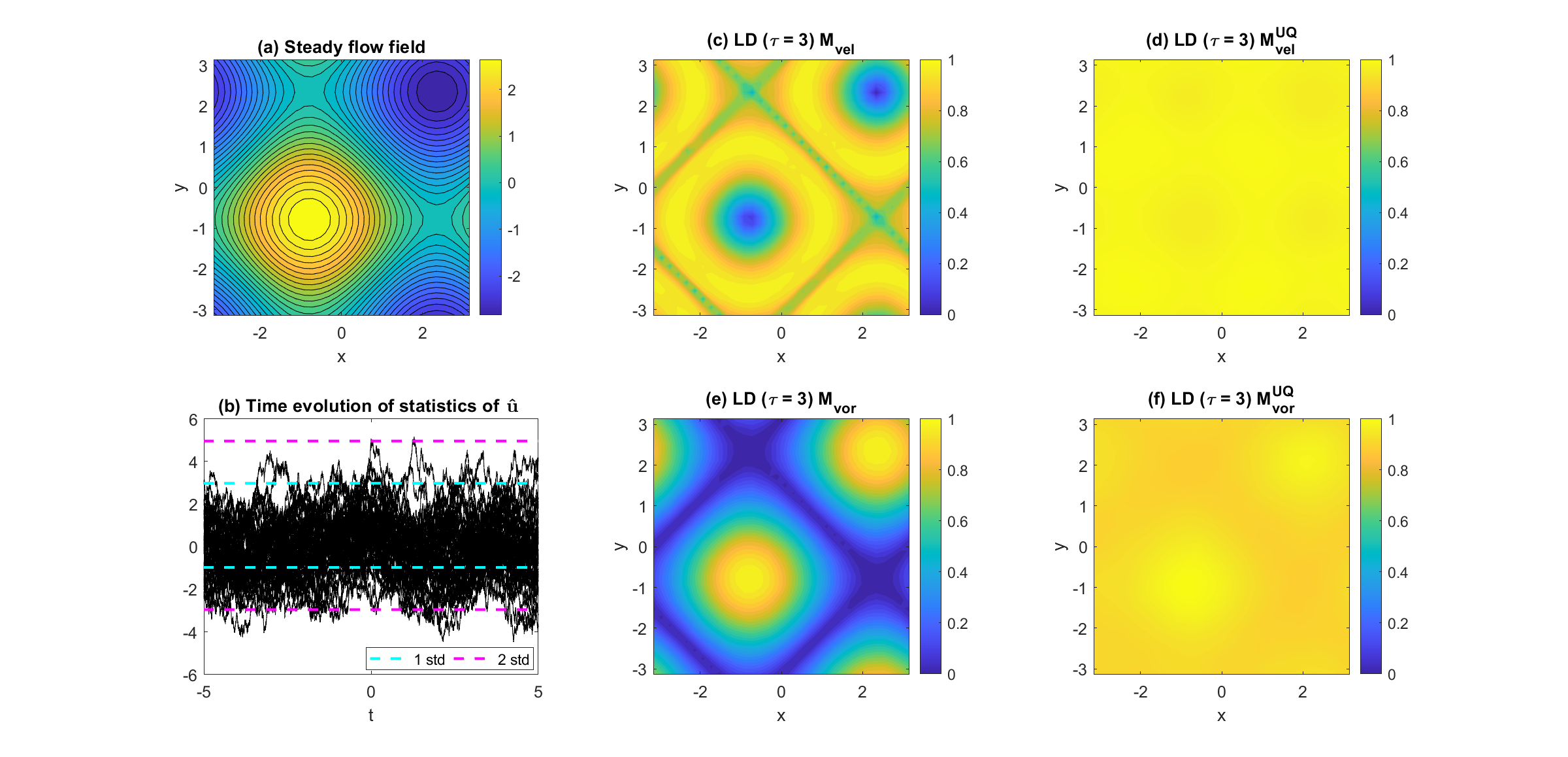}
  \caption{The Lagrangian descriptors for the flow field with the dipole structure. Panel (a): The true steady flow field using stream lines satisfied by \eqref{Dipole_flow} with $\hat{u}\equiv1+i$ and $\hat{v}\equiv1+i$. Panel (b): The statistical equilibrium state of the real part of $\hat{u}$ (dashed lines) and $N_{MC}=50$ sampled trajectories (solid black curves). Panel (c): Velocity-based Lagrangian descriptor (LD) $M_{vel}$ using \eqref{LD_VelocityBased}. Panel (d): Vorticity-based Lagrangian descriptor $M_{vor}$ using \eqref{LD_VorticityBased}. Panel (e): Velocity-based Lagrangian descriptor (LD) with uncertainty $M^{UQ}_{vel}$ using \eqref{LD_VelocityBased_UQ}. Panel (f): Vorticity-based Lagrangian descriptor (LD) with uncertainty $M^{UQ}_{vor}$ using \eqref{LD_VorticityBased_UQ}. In computing all the Lagrangian descriptors, the value in determining the bounds of the integration is chosen as $\tau=3$. }\label{fig: dipole_LD}
\end{figure}

The second flow field is given by a summation of two Fourier modes,  where the stream function $\psi$ and the two velocity components $(u, v)$ are given by
\begin{equation}\label{Dipole_flow}
\begin{aligned}
  \psi(x,y,t) &= \frac{1}{2}\left(\hat{u}(t)e^{iy} + \hat{v}(t)e^{ix} + c.c.\right),\\
   u(x,y,t) &= -\frac{1}{2}i\hat{u}(t)e^{iy} + c.c.,\qquad\mbox{and}\\
   v(x,y,t) &= \frac{1}{2}i\hat{v}(t)e^{ix} + c.c.,
\end{aligned}
\end{equation}
respectively, where c.c. means the complex conjugate. For the true flow field, the coefficients $\hat{u}$ and $\hat{v}$ are assumed to be constants $\hat{u}(t)\equiv 1+i$ and $\hat{v}(t)\equiv 1+i$. See Panel (a) of Figure \ref{fig: dipole_LD}. It has a dipole structure when $\hat{u}$ and $\hat{v}$ have comparable amplitudes (as in the true flow field) but may display different structures when one of the coefficients is significantly larger than the other \cite{majda2006nonlinear, chen2016model}. Similar to the first case, uncertainty is introduced by assuming $\hat{u}$ and $\hat{v}$ are not perfectly known. They are driven by two independent complex-valued OU processes:
\begin{equation}\label{Dipole_SDE}
\begin{aligned}
  \frac{\d\hat{u}}{\d t} &= (-d_u + i\omega_u)\hat{u} + f_u + \sigma_u\dot{W}_u,\\
  \frac{\d\hat{v}}{\d t} &= (-d_v + i\omega_v)\hat{v} + f_v + \sigma_v\dot{W}_v.
\end{aligned}
\end{equation}
Here the parameters are given by
\begin{equation}\label{Dipole_parameters}
\begin{aligned}
  &d_u = 0.8,\qquad \omega_u = 1.0,\qquad  \sigma_u = 2.5,\qquad\mbox{and}\qquad f_u = 0.8(1+i),\\
  &d_v = 0.5,\qquad \omega_v = 0.5,\qquad  \sigma_v = 1.5,\qquad\mbox{and}\qquad f_v = 0.5(1+i),\\
\end{aligned}
\end{equation}
respectively.

Due to the underlying flow structures, the two flow fields are named the quadrupole and dipole flow cases. The Lagrangian descriptor is computed within the interval $[t^*-\tau, t^*+\tau]$, where $t^*=0$. At time $t^*=0$, the PDF of $\hat{u}$ is assumed to be at its equilibrium with mean $f_u/d_u=1$ and variance $\sigma_u^2/(2d_u)=6.25$. Therefore, the statistics of $\hat{u}$ remain constant in time within the interval $[t^*-\tau, t^*+\tau]$. In total, $N_{MC}=50$ time series of $\hat{u}(t)$ are drawn. At each time instant, the distribution of $\hat{u}$ is given by the equilibrium distribution. The time series also has a temporal memory that depends on the damping coefficient $d_u$. These time series are generated by running the governing equation \eqref{Quadrupole_SDE} forward and backward in time with an initial value at $t^*=0$ drawn from the equilibrium distribution. See Panel (b) of Figure \ref{fig: quadrupole_LD}. Similarly, the pair of time series $(\hat{u}(t),\hat{v}(t))$ in the dipole case is drawn following the equations \eqref{Dipole_SDE} and the resulting $N_{MC}=50$ time series of $\hat{u}(t)$ are displayed in Panel (b) of Figure \ref{fig: dipole_LD}.

Panels (c) and (e) in Figure \ref{fig: quadrupole_LD} show the Lagrangian descriptor based on the deterministic true flow field with $\hat{u}\equiv 1$ for the quadrupole case. These two panels display the field velocity-based and the vorticity-base Lagrangian descriptors, namely $M_{vel}$ in \eqref{LD_VelocityBased} and $M_{vor}$ in \eqref{LD_VorticityBased}, respectively. The value $\tau=5$ is used for the integration. The two Lagrangian descriptors reveal geometric structures of the flow field from different aspects. The velocity-based Lagrangian descriptor $M_{vel}$ indicates the structure associated with the transportation of the flow particles. The flow is nearly static at the centers of the vortices $(\pm\pi/2,\pm\pi/2)$ and their intersection $(0,0)$. In contrast, the vorticity-based Lagrangian descriptor $M_{vor}$ is more appropriate for identifying the structure of the vortices and eddies. Different from the low values at the center of the vortices from $M_{vel}$, the highest value of $M_{vor}$ appears at these centers, indicating the strongest locations of the vortices. Panels (d) and (f) show the Lagrangian descriptors with uncertainty, namely $M_{vel}^{UQ}$ in \eqref{LD_VelocityBased_UQ} and $M_{vor}^{UQ}$ in \eqref{LD_VorticityBased_UQ}. These Lagrangian descriptors do not show significant differences from their counterparts using the deterministic flows shown in Panels (c) and (e). Such a result seems to be counterintuitive, especially given the factor that the uncertainty in $\hat{u}(t)$ is quite significant, where the standard deviation $\mbox{std}(\hat{u})=2.5$ is higher than the mean value $\overline{\hat{u}}=1$. The qualitative conclusion holds even with a further increase of $\tau$. The reason will be explained shortly when compared with the dipole case.

Panels (c)--(f) in Figure \ref{fig: dipole_LD} show the Lagrangian descriptors for the dipole case with $\tau=3$. Fundamentally different from the quadrupole case in Figure \ref{fig: quadrupole_LD}, the velocity-based Lagrangian descriptor $M_{vel}^{UQ}$ is homogeneous in the domain, and the geometric structure becomes wholly blurred in the presence of uncertainty. The vorticity-based Lagrangian descriptor $M_{vel}^{UQ}$ has a similar behavior. The structures vaguely seen in $M_{vel}^{UQ}$ will disappear when $\tau$ is further increased to, for example, $\tau=5$.

\begin{figure}[htb]\centering
  \hspace*{-1cm}\includegraphics[width=16cm]{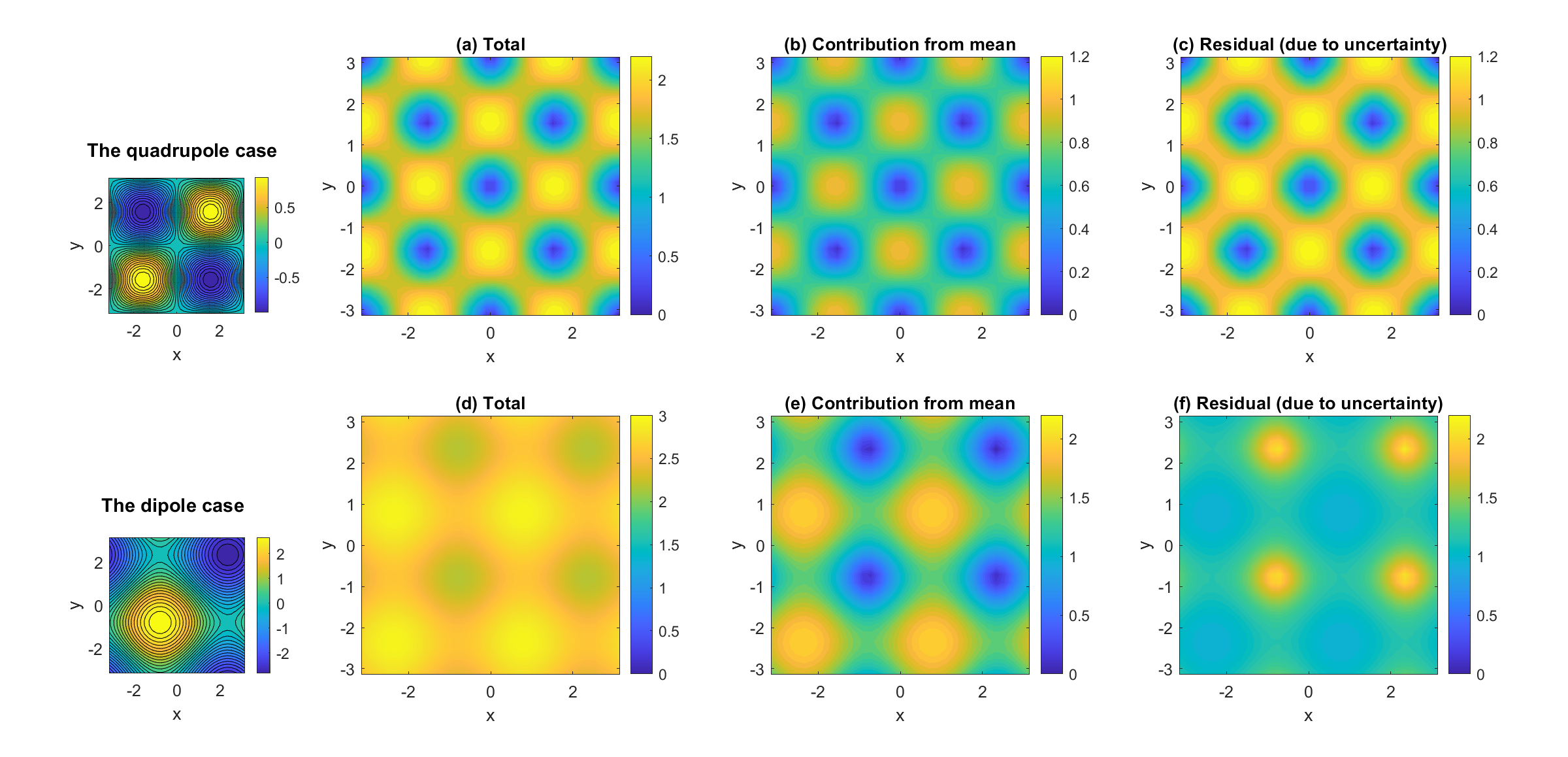}
  \caption{The spatial fields of the positive quantity $\mathcal{F}_E(\mathbf{x},t) = \mathbb{E}_{\mathbf{u}}\left[\sqrt{u(\mathbf{x},t)^2+v(\mathbf{x},t)^2}\right]$ in \eqref{LD_VelocityBased_UQ_2_partI} associated with the velocity-based Lagrangian descriptor at a fixed time instant. Panels (a) and (d): The spatial field of $\mathcal{F}_E(\mathbf{x},t)$ (Total). Panels (b) and (e): The spatial fields constructed by only the contribution from the mean $\sqrt{\overline{u}(\mathbf{x},t)^2+\overline{v}(\mathbf{x},t)^2}$, which are also the results from the deterministic Lagrangian descriptor. Panels (c) and (f): The fields of the resulting uncertainty, which are defined by the difference between the total and the mean contribution. The top three rows show the results of the quadrupole case, while the bottom three rows show those of the dipole case. To illustrate the amplitude of different components, the Lagrangian descriptor and its components are not normalized to the maximum values. }\label{fig: quadrupole_dipole_analysis}
\end{figure}

Figure \ref{fig: quadrupole_dipole_analysis} explains the distinguished behavior of the two flow cases. It shows the spatial fields of the positive quantity $\mathcal{F}_E(\mathbf{x},t) = \mathbb{E}_{\mathbf{u}}\left[\sqrt{u(\mathbf{x},t)^2+v(\mathbf{x},t)^2}\right]$ in \eqref{LD_VelocityBased_UQ_2_partI} associated with the velocity-based Lagrangian descriptor at a fixed time instant. The top three rows show the results of the quadrupole case, while the bottom three rows show those of the dipole case. Panels (a) and (d) show the total spatial field of $\mathcal{F}_E(\mathbf{x},t)$, which is the integrand in computing the Lagrangian descriptor. Panels (b) and (e) show the spatial fields constructed by only the contribution from the mean $\sqrt{\overline{u}(\mathbf{x},t)^2+\overline{v}(\mathbf{x},t)^2}$. As the true velocity equals the mean velocity in both cases, these panels provide the spatial fields resulting from the deterministic Lagrangian descriptor.  Panels (c) and (f) show the residual, which is defined by the difference between the total and the mean contribution. It results from incorporating the uncertainty into the calculation of the Lagrangian descriptor.
The following conclusions can be drawn from this figure. First, it can be computed from the model in the quadrupole case that
\begin{equation}\label{Quadrupole_velocity_sqrt}
  \mbox{var}\left(\sqrt{u^2(x,y,t)+v^2(x,y,t)}\right) =\mbox{var}(\hat{u}(t))\sqrt{\big(\sin(x)\cos(y)+\cos(x)\sin(y)\big)}.
\end{equation}
The variance of $\sqrt{u^2(x,y,t)+v^2(x,y,t)}$ is spatially dependent. As the variance is associated with the uncertainty in $\mathcal{F}_E(\mathbf{x},t)$, the uncertainty is also a function of $(x,y)$. Note that the variance is zero when $x$ and $y$ take values at $0,\pm\pi$, or one of them equals $0,\pm\pi$, but the other is $\pm\pi/2$. These are also the locations where the velocity equals zero; the Lagrangian trajectories have zero arc lengths. The Lagrangian descriptor remains zero when these points are chosen as $\mathbf{x}^*$. Such a finding is consistent with the numerical simulation in Panel (c). Therefore, the uncertainty is strongly inhomogeneous at different locations. According to Panels (b)--(c), the uncertainty of $\mathcal{F}_E(\mathbf{x},t)$ is mostly proportional to its value when the mean velocity is used, i.e., the deterministic situation. This explains why the Lagrangian descriptors with and without taking into account the uncertainty in Panels (c)--(d) of \ref{fig: quadrupole_dipole_analysis} resemble each other. Next, the variance of the arc length in the dipole case has a different spatial pattern. In light of \eqref{Dipole_flow}, it can be seen that
\begin{equation}\label{dipole_velocity_sqrt}
  \mbox{var}\left(\sqrt{u^2(x,y,t)+v^2(x,y,t)}\right) = 2\mbox{var}(\hat{u}(t)) + 2\mbox{var}(\hat{v}(t)),
\end{equation}
which has no spatial dependence. Although the variance is only one of the contributors to the spatially inhomogeneous distribution of the uncertainty (Panel (f) of Figure \ref{fig: quadrupole_dipole_analysis}), it leads to the feature that the uncertainty is nonzero everywhere in the field. According to Panels (e)--(f), the contribution from the mean velocity $\sqrt{\overline{u}(\mathbf{x},t)^2+\overline{v}(\mathbf{x},t)^2}$ and the uncertainty have opposite spatial patterns with comparable amplitudes. This leads to the overall Lagrangian descriptor being a nearly spatial homogenous field (Panel (d)).

\begin{figure}[htb]\centering
  \hspace*{-1cm}\includegraphics[width=16cm]{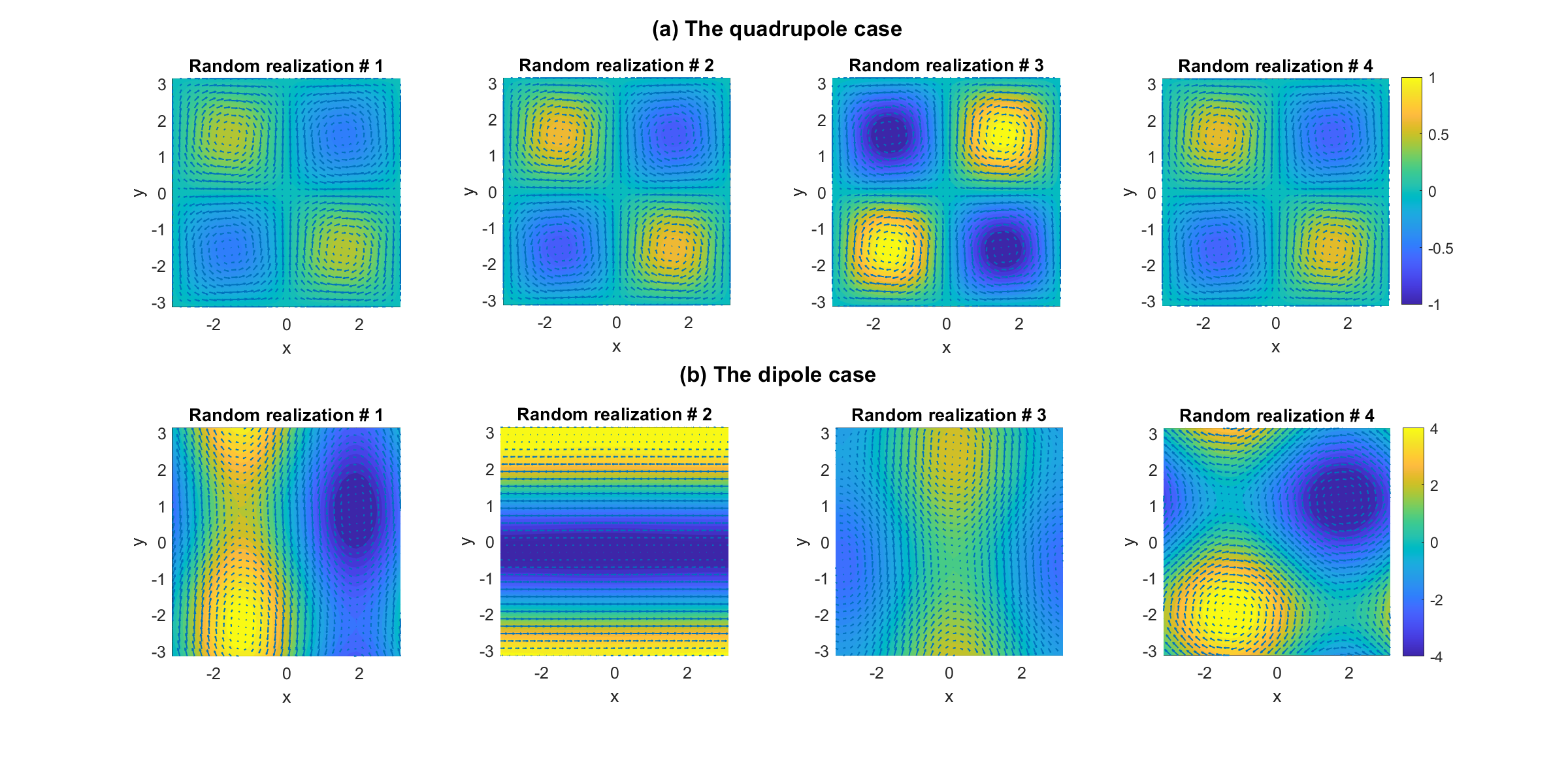}
  \caption{Four random realizations of the flow fields at a fixed time instant when the Fourier coefficients contain uncertainties. Top: The quadrupole case. Bottom: The dipole case. The color bars of the four panels in each row have been scaled to be the same. }\label{fig: quadrupole_dipole_examples}
\end{figure}

Figure \ref{fig: quadrupole_dipole_examples} demonstrates additional evidence of the above finding. It includes four random realizations at a fixed time instant in each case, corresponding to the use of random time series of $\hat{u}$ in the quadrupole case and $\hat{u}$ and $\hat{v}$ in the dipole case. Notably, the randomness only causes the differences in the amplitude of the quadrupole case while the flow structure remains the same. Therefore, the Lagrangian descriptor remains to provide a clear spatial pattern even in the presence of uncertainty. In contrast, the randomness in $\hat{u}$ and $\hat{v}$ breaks the balance of these two values. Consequently, the flows exhibit jet structures, or the locations of the dipoles are shifted when the amplitudes of $\hat{u}$ and $\hat{v}$ become significantly different. Therefore, the uncertainty breaks the original deterministic geometric structure.

\begin{figure}[htb]\centering
  \hspace*{-1cm}\includegraphics[width=16cm]{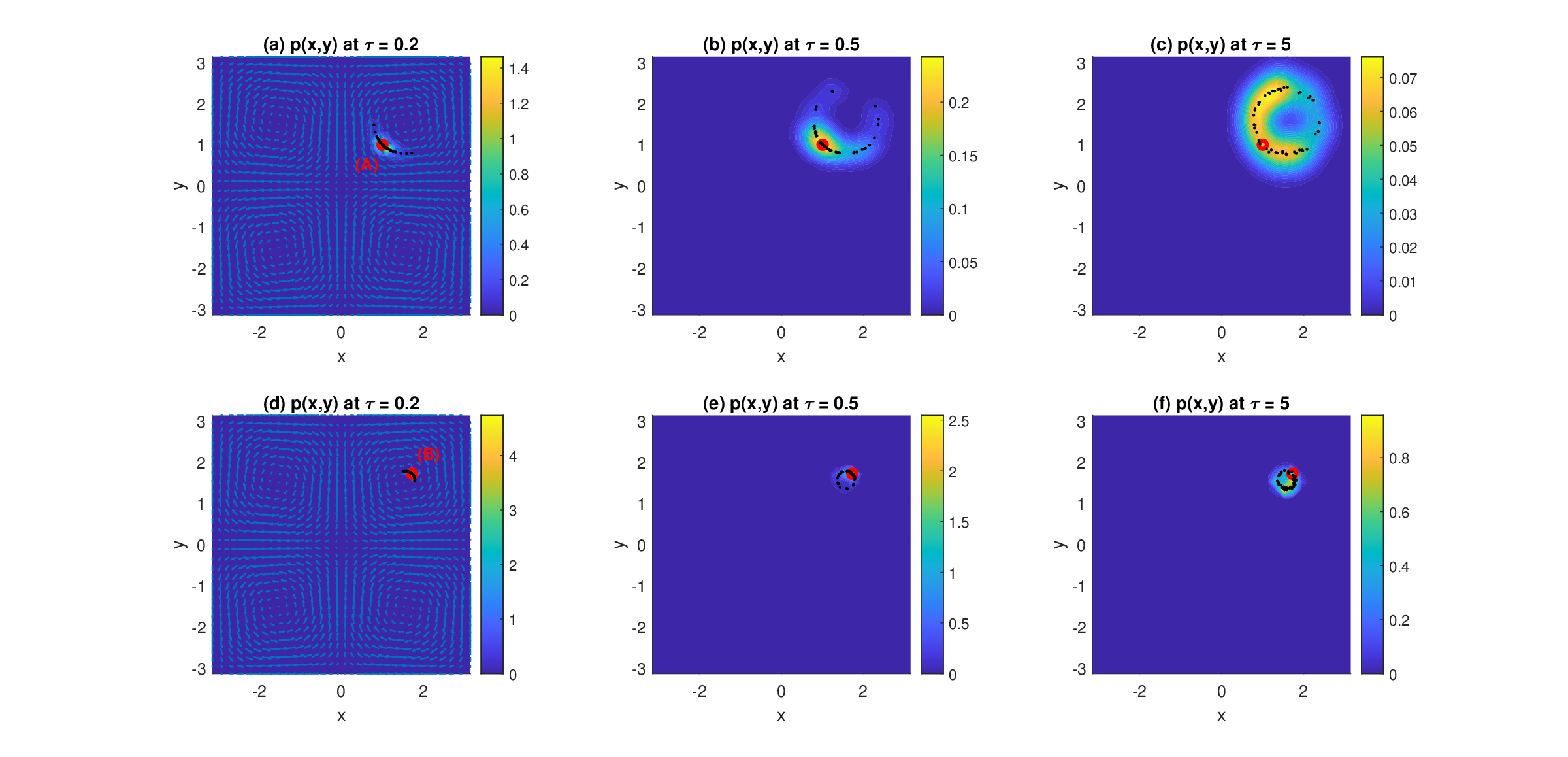}
  \caption{Probability density function (PDF) $p(\mathbf{x})$ for the quadrupole case at time $t^*+\tau$, where $\tau$ takes values of $\tau=0.2$ (Panels (a) and (d)), $\tau=0.5$ (Panels (b) and (e)) and $\tau=5$ (Panels (c) and (f)). The red dot in each panel is the location at time $t^*$, the center of the integration in computing the Lagrangian descriptor. The black dots are the $N_{MC}=50$ ensemble members ending at the time $t^*+\tau$. The three panels in the top row have the same starting point, $A$. So do the three ones in the bottom row with the point denoted by $B$. Point $A$ is between the edge and the center of the vortex, while point $B$ is near the center of the vortex. The true flow field is shown in Panels (a) and (d) for reference, but the actual flow field will have different structures when uncertainty comes, as shown in Figure \ref{fig: quadrupole_dipole_examples}. }\label{fig: quadrupole_PDF}
\end{figure}

\begin{figure}[htb]\centering
  \hspace*{-1cm}\includegraphics[width=16cm]{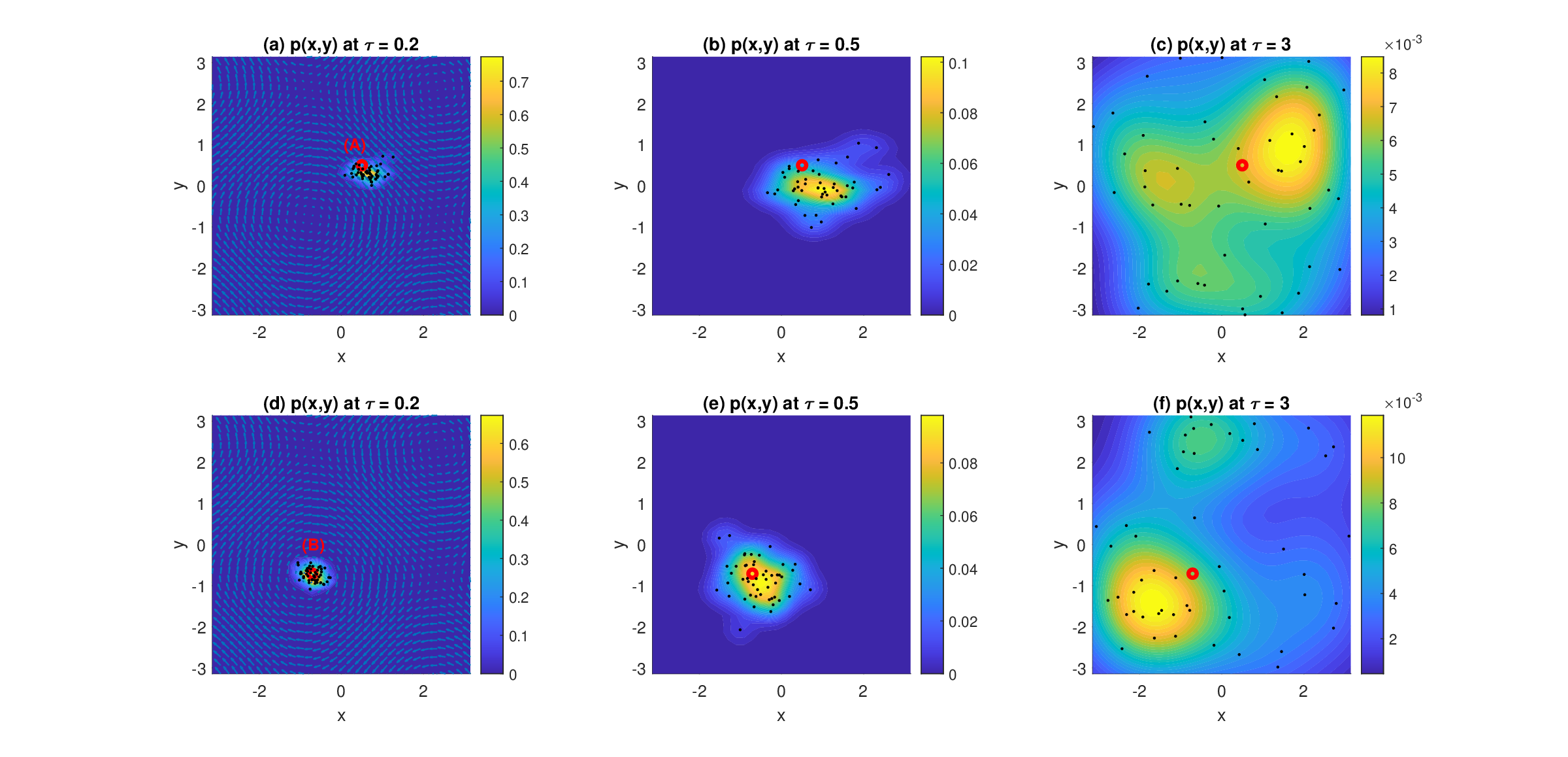}
  \caption{Probability density function (PDF) $p(\mathbf{x})$ for the dipole case at time $t^*+\tau$, where $\tau$ takes values of $\tau=0.2$ (Panels (a) and (d)), $\tau=0.5$ (Panels (b) and (e)) and $\tau=3$ (Panels (c) and (f)). The red dot in each panel is the location at time $t^*$, the center of the integration in computing the Lagrangian descriptor. The black dots are the $N_{MC}=50$ ensemble members ending at the time $t^*+\tau$. The three panels in the top row have the same starting point, $A$. So do the three ones in the bottom row with the point denoted by $B$. Point $A$ is at the boundary between the two vortices, which has the fastest flow velocity, while point $B$ is closer to the center of one of the vortices. The true flow field is shown in Panels (a) and (d) for reference, but the actual flow field will have different structures when uncertainty comes, as shown in Figure \ref{fig: quadrupole_dipole_examples}. }\label{fig: dipole_PDF}
\end{figure}

Figures \ref{fig: quadrupole_PDF}--\ref{fig: dipole_PDF} illustrate the PDF $p(\mathbf{x})$ for the quadrupole and the dipole case, respectively, at time $t^*+\tau$. As is expected, the ensemble spreads increase when $\tau$ becomes large. Because the flow structure remains in the same profile in the quadrupole case, the ensemble members follow the circle inside the vortex near their initial values. At different locations within the circle, the value $\sqrt{u^2+v^2}$ remains the same. In contrast, as the flow structure becomes very different due to the randomness, the ensemble members are located randomly in the entire field. Consequently, regardless of the initial location $\mathbf{x}^*$, the Lagrangian descriptor tends to average over the flow information at all grids. Thus, the spatial averaging of the information highly erodes the resulting geometric structure and becomes nearly homogeneous.

To summarize, in the presence of uncertainty, the geometric structure provided by the Lagrangian descriptor may become significantly different from the one based on the deterministic flow field. The uncertainty can completely erode the coherent structure and leads to a fully noninformative geometric pattern. However, a large uncertainty does not always mean the inferred geometric structure will become blurred. The uncertainty may not necessarily have a significant impact on the geometric structure of the underlying flow. The way the uncertainty interacts with the deterministic patterns depends on the flows. Therefore, incorporating the uncertainty in computing the Lagrangian descriptor is crucial in understanding if the uncertainty plays a central role in disturbing the geometric structure.

\subsection{Eddy detection with velocity field recovered from Lagrangian data assimilation}
Ocean eddy detection is an important topic that helps in understanding the transport and mixing of fluid and its impact on ocean features. Lagrangian descriptors have been used in eddy detection and have been shown to outperform the Eulerian-based methods and some other trajectory diagnostic approaches \cite{vortmeyer2016detecting, branicki2011lagrangian, vortmeyer2019comparing}. This subsection focuses on studying the eddy detection skill using Lagrangian descriptors in the presence of uncertainty when the ocean velocity field is recovered from Lagrangian data assimilation.

\begin{figure}[htb]\centering
  \hspace*{-1cm}\includegraphics[width=16cm]{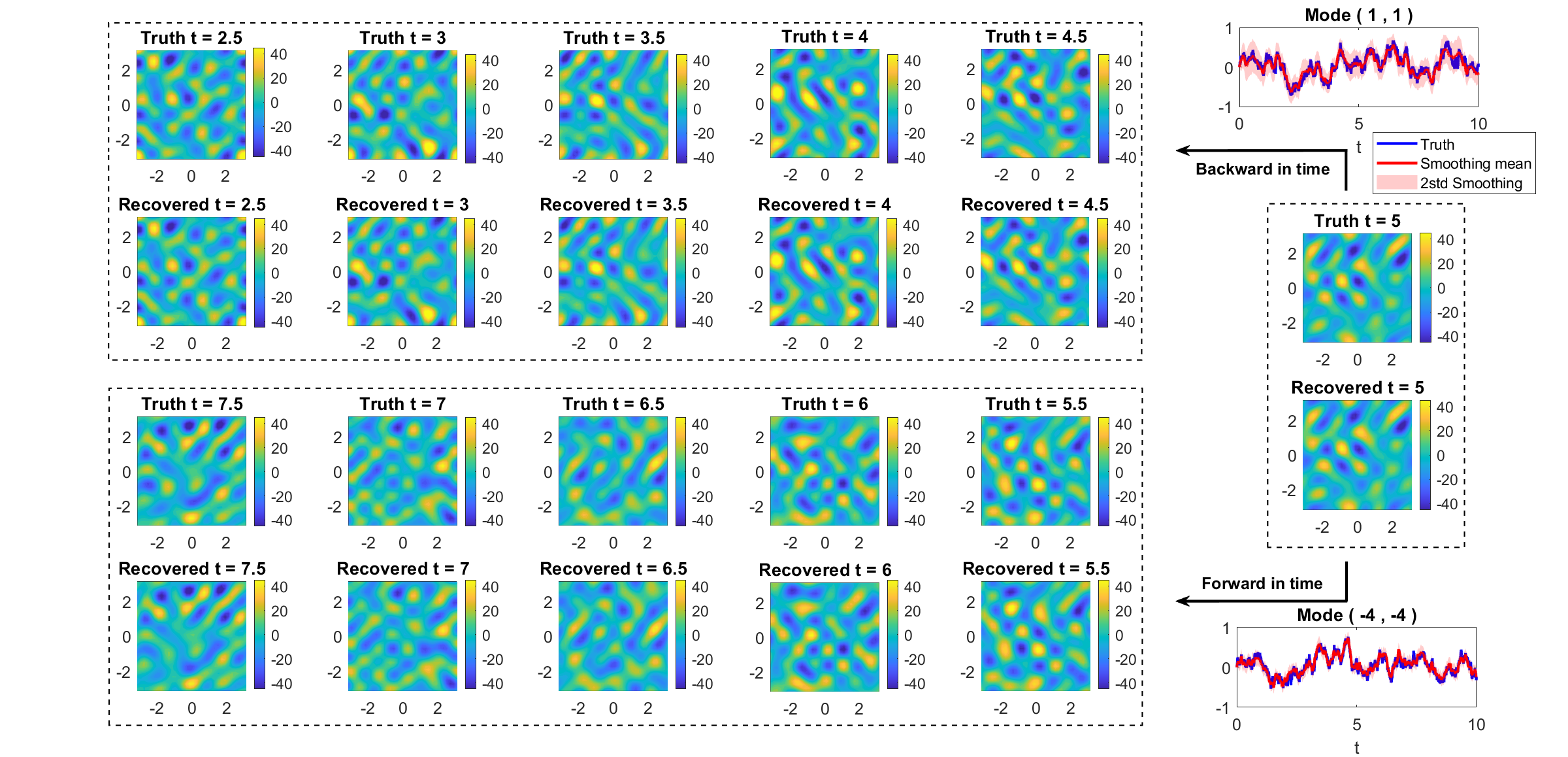}
  \caption{Solution of the flow field associated with the eddy detection problem. The contour plots with shadings show the vorticity fields at different time instants. The two panels on the top right and bottom right show the true series of Fourier modes $(1,1)$ and $(-4,-4)$ (blue) and the data assimilation solution, where the posterior mean is presented by the solid red curve and the confidence interval is shown in the red shading area. }\label{fig: Vorticity_all_time}
\end{figure}

The ocean model considered here is given by \eqref{Ocean_Velocity}--\eqref{OU_process} with double periodic boundary conditions, where $K_{\mbox{max}}=4$ is utilized. The flow field is assumed to be incompressible without any mean background flow. Therefore, there are in total of $80$ Fourier modes. The parameters are
\begin{equation}\label{Parameters_eddy_model}
  d_\mathbf{k} = 1,\qquad \omega_\mathbf{k}=0,\qquad f_\mathbf{k}=0\qquad \mbox{and}\qquad \sigma_\mathbf{k} = 0.75.
\end{equation}
for all $\mathbf{k}$. Therefore, the flow field has an equipartition of the energy. In total, $L=32$ Lagrangian tracers are used to recover the underlying flow field. The initial distribution of tracers is uniform, which is consistent with the statistical equilibrium state \cite{chen2014information}. The time instant $t^*=5$ is chosen for eddy detection. Figure \ref{fig: Vorticity_all_time} shows the vorticity fields at different time instants. It also compares the true vorticity field with the recovered field based on the posterior mean. They resemble each other. The level of uncertainty is moderate and appears mainly in the large-scale modes, as can be seen in the red shading area in the top right panel in Figure \ref{fig: Vorticity_all_time}. Note that only the diagonal entries of the matrix $\mathbf{R}$ in \eqref{eq:filter} are saved and applied to computing the smoother and sampling solutions \eqref{Smoother_Main}--\eqref{Sampling_Main}. This significantly reduces computational storage and introduces little error. In fact, when the flow field is incompressible, it has been shown that $\mathbf{R}$ will converge to a diagonal matrix when $L$ increases \cite{chen2014information}.

\begin{figure}[htb]\centering
  \hspace*{-1cm}\includegraphics[width=16cm]{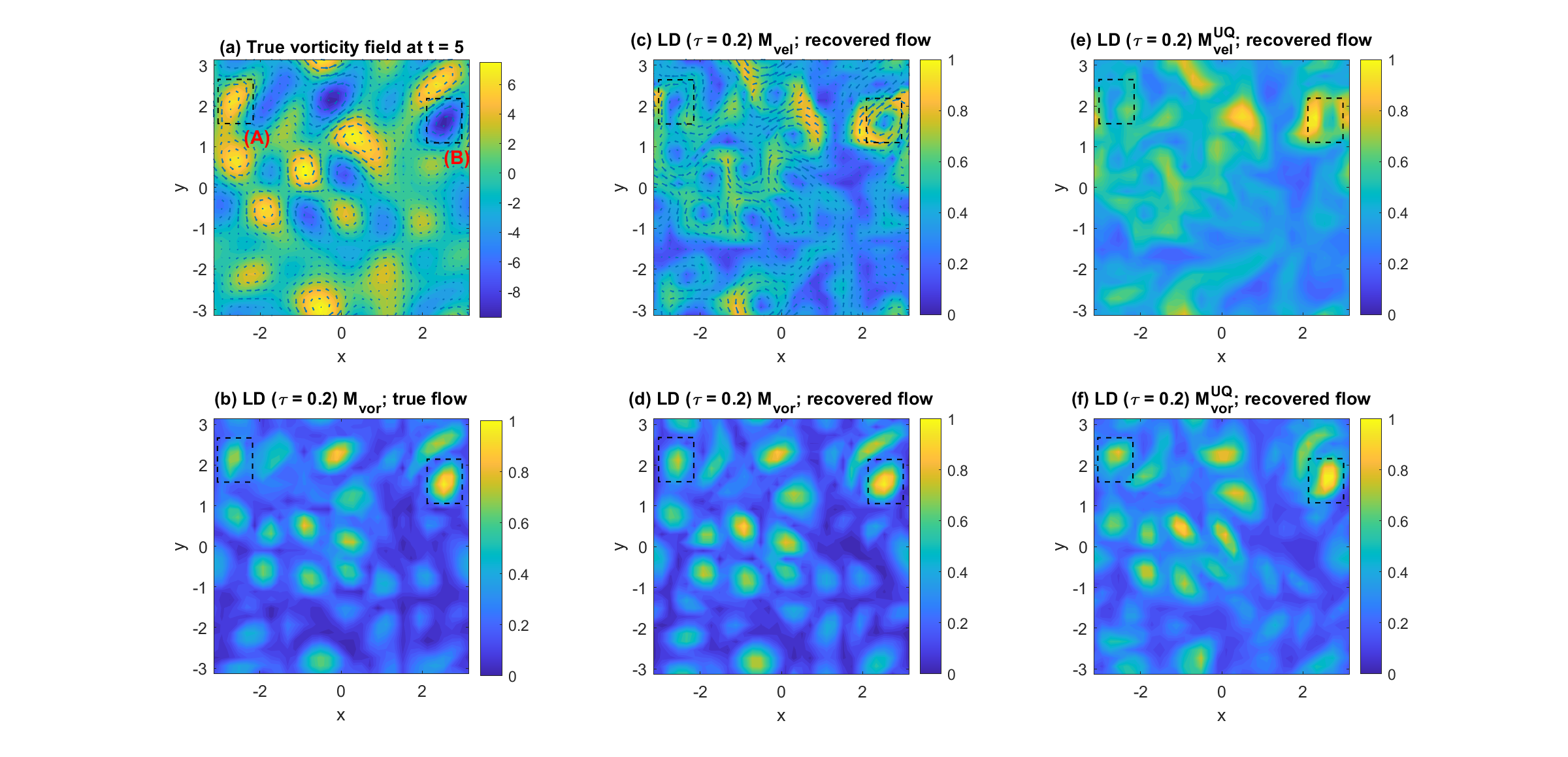}
  \caption{Eddy detection using the Lagrangian descriptors. Panel (a): The true vorticity field at $t=5$. Panel (b): The vorticity-based Lagrangian descriptor using the true signal. Panels (c)--(d): The velocity-based and the vorticity-based Lagrangian descriptors using the recovered flow field from data assimilation, where only the posterior mean time series is used. Panels (e)--(f): The velocity-based and the vorticity-based Lagrangian descriptors using the recovered flow field from data assimilation, where the entire posterior distribution is used, and therefore the uncertainty is considered. The arrows in Panel (a) show the true velocity field, while those in Panel (b) show the recovered one based on the posterior mean. The black dashed boxes mark two eddies, named (A) and (B), discussed in the main text. In all the Lagrangian descriptors computed here, $\tau=0.2$ is utilized.}\label{fig: Vorticity_identify_tau02}
\end{figure}

\begin{figure}[htb]\centering
  \hspace*{-1cm}\includegraphics[width=16cm]{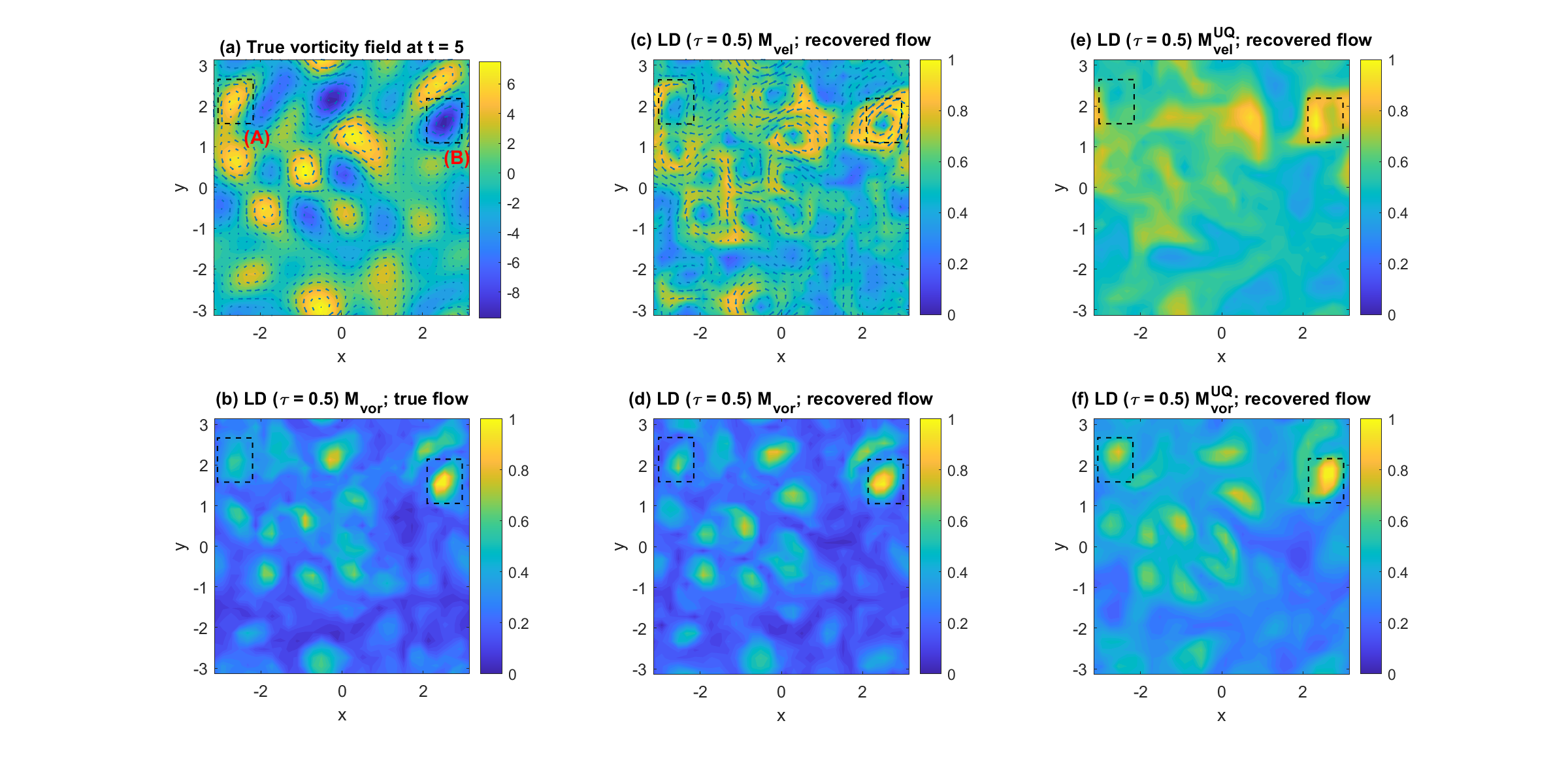}
  \caption{Similar to Figure \ref{fig: Vorticity_identify_tau02}, but with $\tau = 0.5$. }\label{fig: Vorticity_identify_tau05}
\end{figure}

\begin{figure}[htb]\centering
  \hspace*{-1cm}\includegraphics[width=16cm]{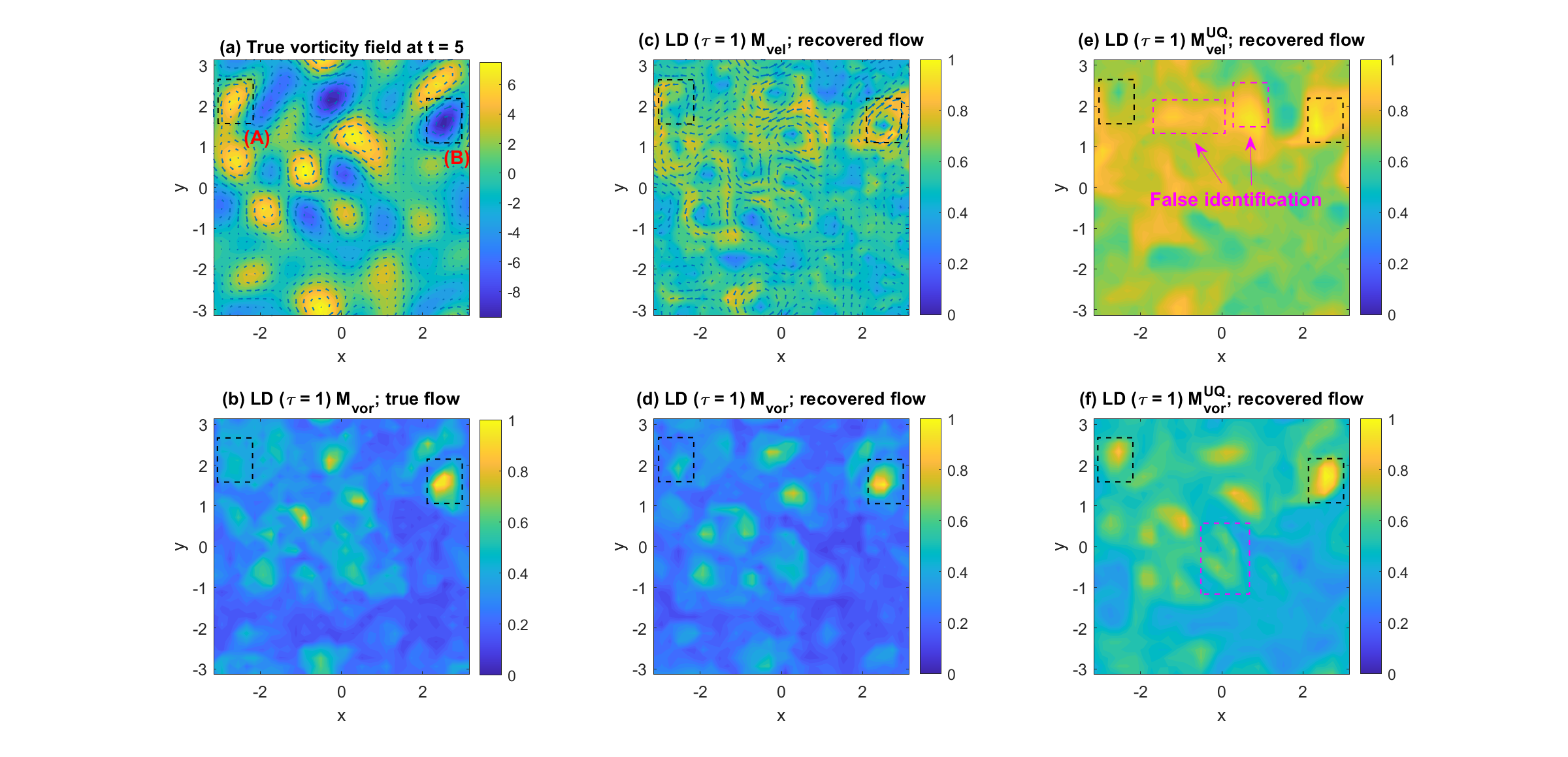}
  \caption{Similar to Figure \ref{fig: Vorticity_identify_tau02}, but with $\tau = 1$. }\label{fig: Vorticity_identify_tau1}
\end{figure}

Different from revealing the stationary structure of the dynamical systems in many other applications, the choice of a suitable size of the integration window in computing the Lagrangian descriptor, namely $\tau$ in \eqref{LD_VelocityBased_UQ} or \eqref{LD_VorticityBased_UQ}, is crucial for eddy detection. This is because eddies occur intermittently and are not a feature that is reflected in the long-term average. Figures \ref{fig: Vorticity_identify_tau02}--\ref{fig: Vorticity_identify_tau1} show the eddy detection results with $\tau=0.2$, $\tau=0.5$ and $\tau=1$, respectively. Note that the decorrelation times of different Fourier modes are all $1/d_\mathbf{k}=1$, which can also be roughly regarded as the temporal memory for the entire system and the averaged eddy lifetime. Therefore, the size of the integration window $2\tau$ is small than, equal to, and larger than the decorrelation time of the system, respectively, in the cases shown in these figures. In all these figures, Panel (a) displays the true vorticity field at $t=5$. Panel (b) shows the vorticity-based Lagrangian descriptor using the true signal. Panels (c) and (d) compare the velocity-based and the vorticity-based Lagrangian descriptors \eqref{LD_VelocityBased} and \eqref{LD_VorticityBased} using the recovered flow field from data assimilation, where only the posterior mean time series is used. These are the traditional deterministic Lagrangian descriptors. Panels (e)--(f) show the results when uncertainty is considered in computing the Lagrangian descriptors. The following conclusions can be made from these figures.

First, the vorticity-based Lagrangian descriptor is a more appropriate choice compared with the velocity-based one. Because the arc length of the trajectories trapped in the core of the eddies is short, the velocity-based Lagrangian descriptor leads to small values at the center of the eddies. This raises a fundamental difficulty in eddy detecting because it cannot distinguish the core of eddies with the nearly static flows \cite{vortmeyer2016detecting}. See, for example, the two eddies marked by black dashed boxes in Panel (b) of Figure \ref{fig: Vorticity_identify_tau02}, which have almost the same values as the right bottom part of the domain where the flows have small velocity and no evident eddies are seen. In contrast, the vorticity-based Lagrangian descriptor leads to significant values at the core of eddies, which facilitate eddy detection. In addition to the skill of characterizing eddies, the geometric structure provided by the velocity-based Lagrangian descriptor is also eroded much faster than the vorticity-based Lagrangian descriptor when uncertainty appears. This can be seen by comparing Panels (e) and (f) in Figure \ref{fig: Vorticity_identify_tau1}. The correct locations of eddies at this relatively long time scale can be well identified using the vorticity-based Lagrangian descriptor. As a comparison, the velocity-based Lagrangian descriptor suggests significant values over a large area in the domain. Many of these locations, however, do not contain strong eddies. Note that the Lagrangian descriptor using the posterior mean time series provides a cleaner geometric flow structure than the truth. This is because the truth is one realization from the stochastic (or turbulent) model. The posterior mean time series smooths out small wiggles due to the stochasticity, leading to a less noisy field.

Second, uncertainty mainly affects the eddies over a longer time scale. This is because the Lagrangian trajectories only separate a little within a short time. Therefore the difference between $M_{vor}$ and $M_{vor}^{UQ}$ in Figure \ref{fig: Vorticity_identify_tau02} is much less significant than that in Figure \ref{fig: Vorticity_identify_tau1}.

Third, uncertainty can affect eddy identification in different ways. In many cases, the uncertainty will blur the identified field to give less confidence for the detected eddies, for example, the one marked by the dashed pink box in Panel (f) of Figure \ref{fig: Vorticity_identify_tau1}. This is intuitive as uncertainty is expected to break the coherent structures. Yet, the uncertainty may also impose more confidence on the identified eddies. The eddies marked by the two black boxes in Panel (f) of Figure \ref{fig: Vorticity_identify_tau1} are examples of this category. Eddies, especially the strong ones, can be regarded as local extreme events with significant signals. When the uncertainty induces an overestimation of the signal, it can cause a further enhancement of the eddy amplitude at the core, making the identified eddy more significant. Note that an identified eddy with strong amplitude does not necessarily mean it is more likely to be an actual eddy. The uncertainty can make the actual eddies indistinct or induce fake eddies that do not exist in the true field. This provides a crucial suggestion for eddy detection in practice. That is, instead of concluding the existence or nonexistence of an eddy in a location, assigning a probability to each identified eddy that accounts for the uncertainty is a more appropriate strategy.

\begin{figure}[htb]\centering
  \hspace*{-1cm}\includegraphics[width=16cm]{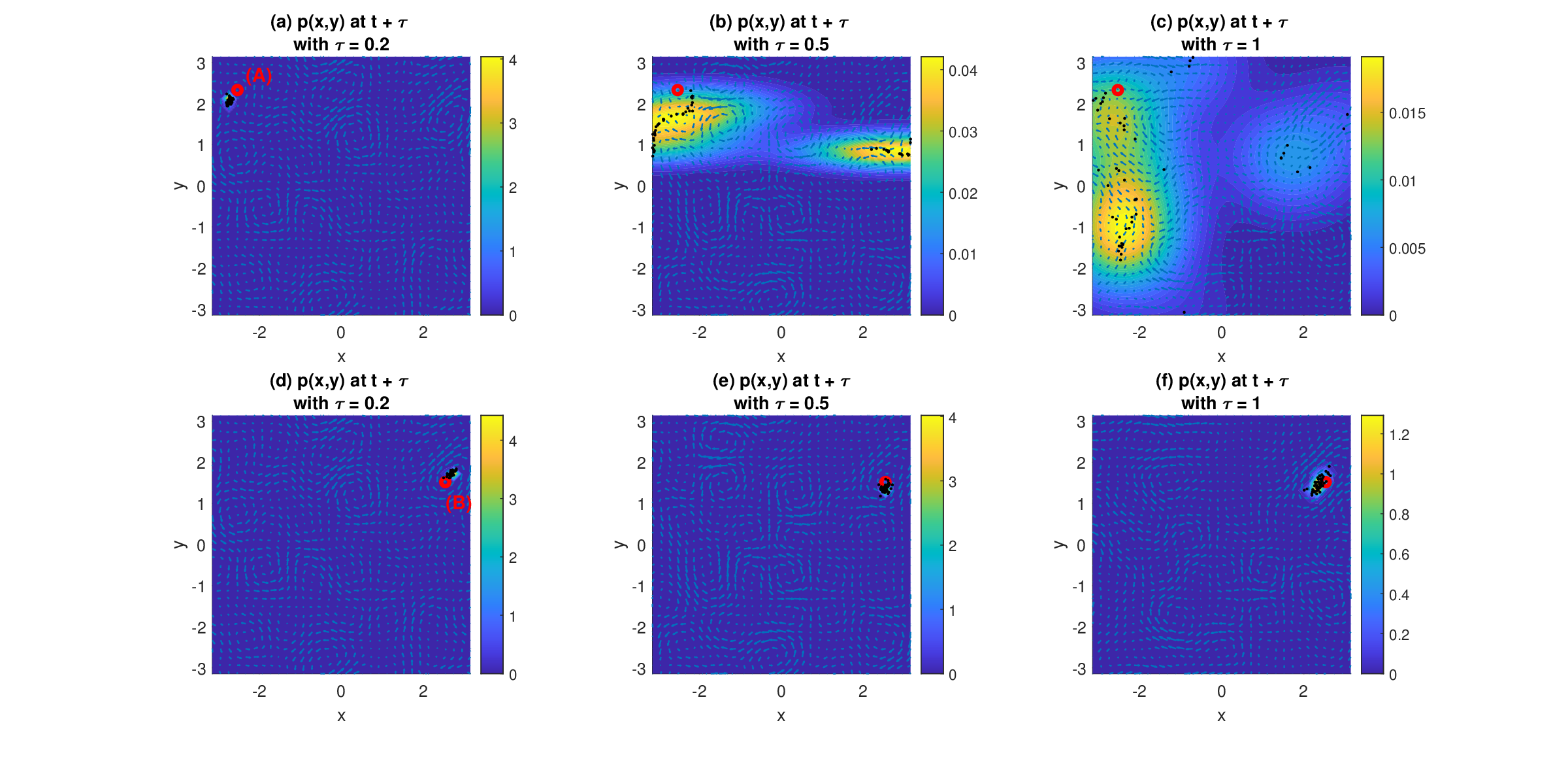}
  \caption{Probability density function (PDF) $p(\mathbf{x})$ for the eddy detection study at a forward time instant $t^*+\tau$, where $\tau$ takes values of $\tau=0.2$ (Panels (a) and (d)), $\tau=0.5$ (Panels (b) and (e)) and $\tau=1$ (Panels (c) and (f)). The red dot in each panel is the location at time $t^*$, the center of the integration in computing the Lagrangian descriptor. The black dots are the $N_{MC}=50$ ensemble members ending at the time $t^*+\tau$. The three panels in the top row have the same starting point, $A$. The three panels in the bottom row have the same starting point, $B$. The arrows in each panel show the true velocity fields at time $t$. }\label{fig: Vorticity_pdf_forward}
\end{figure}

Figure \ref{fig: Vorticity_pdf_forward} shows the PDF $p(\mathbf{x})$ at a forward time instant $t^*+\tau$ starting from the two points (A) and (B) marked in red in Figures \ref{fig: Vorticity_identify_tau02}--\ref{fig: Vorticity_identify_tau1}, which exhibit different behavior. The eddy around point (A) has a strong connection with the one beneath it, which can be seen from, for example, Panel (a) of Figure \ref{fig: Vorticity_identify_tau02}. Such a feature can be identified from the Lagrangian trajectories. When uncertainty arrives, the trajectories diverge to a wide area. In contrast, the eddy around point (B) is a strong local one. The Lagrangian trajectories are all trapped inside the local area that confirms the existence of such a local eddy. As a final remark, the singular lines corresponding to manifolds in the vorticity-based Lagrangian descriptor can systematically provide the shape of eddies. This can be done by searching for the largest closed contour line of the Lagrangian descriptor for which the Lagrangian descriptor is an extremum and which surrounds an eddy core found with the Lagrangian descriptor. The details can be found in \cite{vortmeyer2016detecting}. Such a postprocessing procedure is omitted in Figures \ref{fig: Vorticity_identify_tau02}--\ref{fig: Vorticity_identify_tau1}, as it is not the focus of the current work.

\subsection{Detecting the source target}
Identifying the source target is of practical interest. This subsection aims to reveal the new features in tracing the source when uncertainty appears using the Lagrangian descriptor \eqref{LD_Source_Probability}.

\begin{figure}[!h]\centering
  \hspace*{-1cm}\includegraphics[width=16cm]{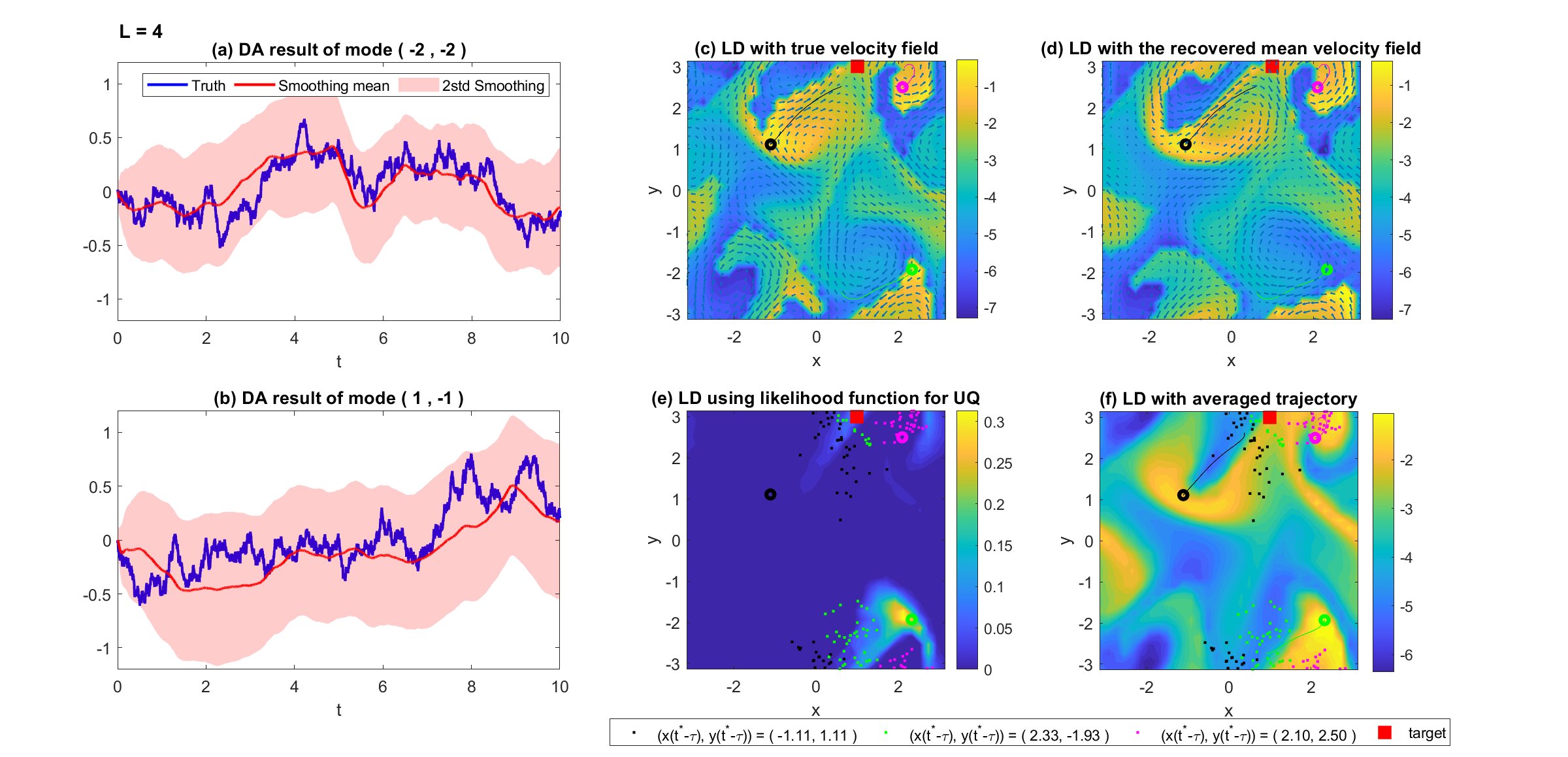}
  \caption{Identifying the source target using the Lagrangian descriptor. The source is one time unit ahead of the target (i.e., $\tau=-1$). Panels (a)--(b): Comparison of the true time series (blue) and the data assimilation results of modes $(-2,-2)$ and $(1,-1)$, where the posterior mean is shown in the red solid curve while the posterior uncertainty (two standard deviations) is shown in the red shading area. The Lagrangian data assimilation results are based on exploiting $L=4$ observed tracer trajectories. Panel (c): The Lagrangian descriptor \eqref{LD_Source_Deterministic} using the true deterministic flow field. Panel (d): The Lagrangian descriptor \eqref{LD_Source_Deterministic} using the posterior mean of the recovered flow field. Panel (e): Panel (f): The averaged value of the Lagrangian descriptor over different random realizations. This is done by first computing the Lagrangian descriptor \eqref{LD_Source_Deterministic} of every random realization of the flow field sampled from the posterior distributions and then taking the average. Note that the contours shown in Panels (c), (d), and (f) are the non-normalized Lagrangian descriptor multiplied by a negative sign. Therefore, yellow shows the most possible sources, consistent with the color bar convention in Panel (e) for the likelihood. In Panels (c)--(f), the red square is the target. The black, pink, and green circles are the three potential sources at $\tau=-1$. The dots are the ensemble forecast at the current time instant starting from each circle, which provides a distribution used to compute the likelihood. }\label{fig: Likelihood_L4}
\end{figure}

\begin{figure}[!h]\centering
  \hspace*{-1cm}\includegraphics[width=16cm]{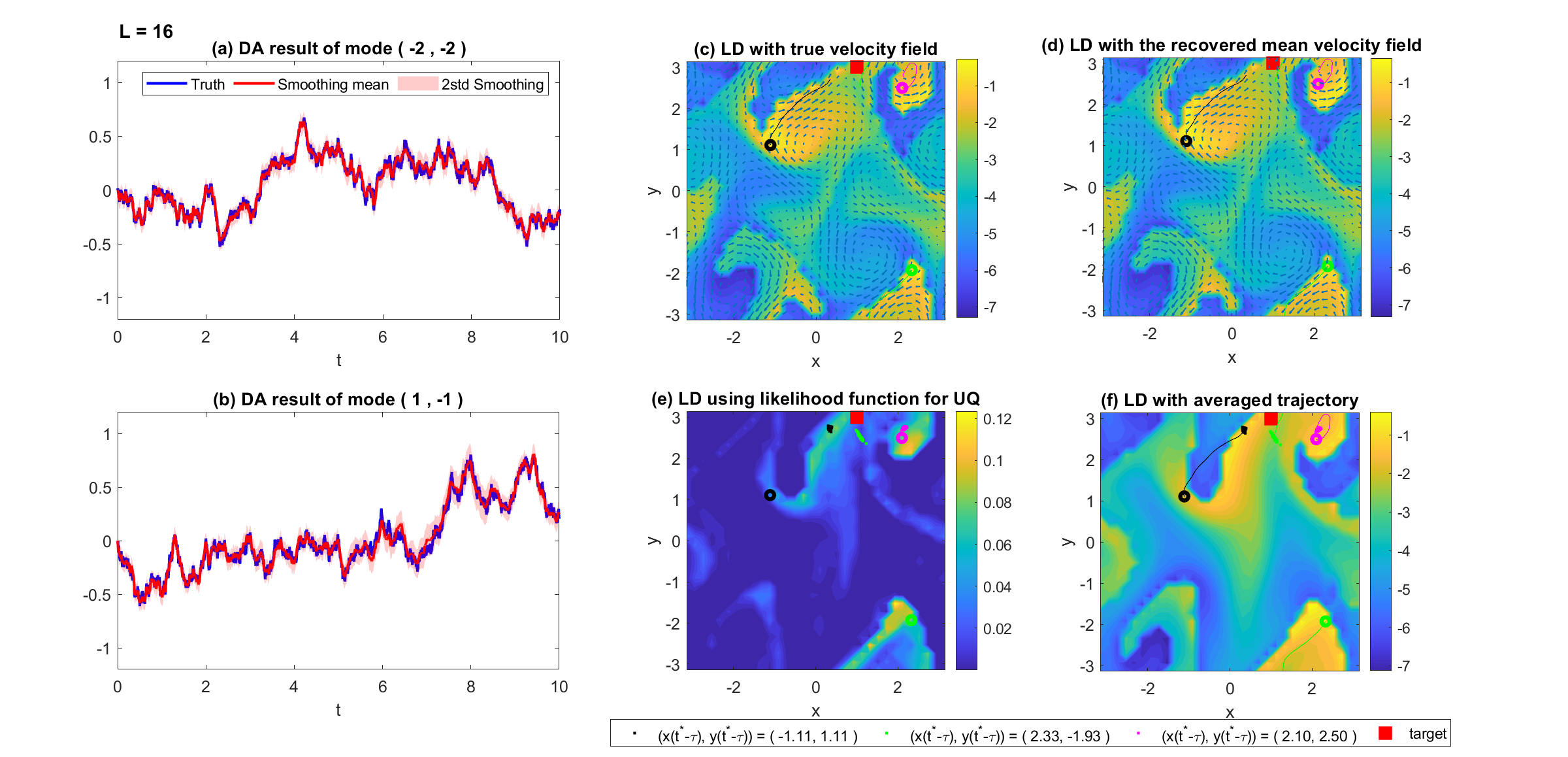}
  \caption{Similar to Figure \ref{fig: Likelihood_L4}, but with $L=16$ tracers in data assimilation. }\label{fig: Likelihood_L16}
\end{figure}

The ocean model considered here is given by \eqref{Ocean_Velocity}--\eqref{OU_process} with double periodic boundary conditions, where $K_{\mbox{max}}=2$ is utilized. The flow field is assumed to be incompressible without any mean background flow. Therefore, there are in total $24$ Fourier modes. The parameters are
\begin{equation}\label{Parameters_eddy_model}
  d_\mathbf{k} = 0.5,\qquad \omega_\mathbf{k}=0,\qquad f_\mathbf{k}=0\qquad \mbox{and}\qquad \sigma_\mathbf{k} = 0.5.
\end{equation}
for all $\mathbf{k}$. The recovered ocean flow field is provided by the Lagrangian data assimilation. Two experiments are carried out with $L=4$ and $L=16$ tracers used in the Lagrangian data assimilation. The corresponding posterior distribution is shown in Panels (a)--(b) of Figure  \ref{fig: Likelihood_L4} and \ref{fig: Likelihood_L16}, respectively. With a relatively small number of tracers, $L=4$, a larger uncertainty is found than with $L=16$.

Panels (c)--(f) of Figure \ref{fig: Likelihood_L4} show the identified sources using different Lagrangian descriptors. Note that in Panels (c) and (d), the contours represent the non-normalized Lagrangian descriptor multiplied by a negative sign. Therefore, yellow implies the most possible sources, consistent with the color bar convention in Panel (e) for the likelihood. Panel (c) shows the Lagrangian descriptor \eqref{LD_Source_Deterministic} using the true deterministic flow field. The three points marked by black, pink, and green circles are the possible sources, starting from which the trajectories will end up at a place close to the target. Panel (d) displays the Lagrangian descriptor using the posterior mean flow field. As the posterior mean captures the truth (Panels (a)--(b)) quite well, the resulting Lagrangian descriptor leads to a similar geometric pattern as that in Panel (c). Unlike the Lagrangian descriptor based on the deterministic flow field, Panel (e) shows the Lagrangian descriptor when the uncertainty is considered using \eqref{LD_Source_Probability}. As the uncertainty has a comparable level as the signal (Panels (a)--(b)), the Lagrangian descriptor in Panel (e) displays a very different pattern. One interesting finding is that only the green circle is now a potential source of the target when uncertainty appears. This is because the PDF represented by the ensemble members (green dots) ``covers'' the target. Therefore, the likelihood of the target is significant. In contrast, the target does not lie inside the resulting PDF inferred from the pink and the black dots, which are the prediction starting from the pink and the black circles. Therefore, the three potential sources have distinguished behavior in the presence of uncertainty. Panel (f) uses the same uncertainty flow field as Panel (e), but the Lagrangian descriptor is computed based on the mean of the forecast ensemble members, which then becomes the deterministic approach \eqref{LD_Source_Deterministic}. This differs from the one in \eqref{LD_Source_Probability} by computing the likelihood using the PDF. The result in Panel (f) is similar to those in Panels (c)--(d). It implies that taking the average of ensemble members before or after forecasting the trajectories has little impact on identifying the geometric structure of the flow. However, computing the Lagrangian descriptor using the full PDF that accounts for the uncertainty will lead to a significantly different result than using only the mean trajectory. Notably, the average distance from two ensembles to the target can be the same. The ensemble spread can provide additional information to distinguish the behavior of the corresponding two starting points.

Figure \ref{fig: Likelihood_L16} is similar to Figure \ref{fig: Likelihood_L4}, but with $L=16$ tracers in data assimilation. Therefore, the uncertainty decreases significantly (Panels (a)--(b)). In such a case, the Lagrangian descriptor based on the likelihood (Panel (e)) leads to a similar result as the deterministic ones (Panels (c), (d), and (f)). This provides numerical evidence that the likelihood-based Lagrangian descriptor is consistent with the standard one in the limit of shrinking the uncertainty.


\section{Conclusion}
This paper develops a mathematical framework for computing Lagrangian descriptors when uncertainty appears. It has the unique feature of characterizing the uncertainties in both the velocity fields and the Lagrangian trajectories. The former is extremely important as it directly affects the crucial nonlinear positive scalar function in computing the Lagrangian descriptor, distinguishing it from existing approaches. The method developed here is also naturally adaptive to the solution from data assimilation, allowing a combination of noisy data with dynamical or statistical models for uncertainty quantification. Simple but illustrative examples are designed to show that uncertainty can either completely erode the coherent structure detected or barely affect the underlying geometry of the flow field using the Lagrangian descriptor. The method has also been applied for idealized eddy identification problems, indicating that uncertainty has distinct impacts on detecting eddy in different time scales. Finally, when uncertainty is incorporated into the Lagrangian descriptor for inferring the source target, the likelihood criterion provides a very different conclusion from the deterministic methods. As a first step, all the tests carried out in this work used synthetic data. Further work includes applying this new tool to practical problems such as referring ocean structures based on floe trackers in the marginal ice zone and tracing the ocean pollution when uncertainty appears in the reanalysis data for the flow field.

\section*{Acknowledgement}
The research of N.C. is funded by ONR N00014-19-1-2421 and ARO W911NF-23-1-0118. The research of E.L. is supported by ONR N0001423WX01622. S~.W.~acknowledges the financial support provided by the EPSRC Grant No. EP/P021123/1 and the support of the William R. Davis '68 Chair in the Department of Mathematics at the United States Naval Academy.

\section{Appendix: Derivation of the results in Proposition \ref{Proposition_1}}
The results in Proposition \ref{Proposition_1} are obtained as followed.
For two positive numbers $A$ and $B$, elementary algebraic calculation shows that
\begin{equation*}
  (\sqrt{A} + \sqrt{B})^2 = A + B + 2 \sqrt{A}\sqrt{B},
\end{equation*}
which gives
\begin{equation*}
  \sqrt{A + B} = \sqrt{(\sqrt{A} + \sqrt{B})^2 - 2 \sqrt{A}\sqrt{B}}.
\end{equation*}
Now let $A$ and $B$ be a positive random variable,
\begin{equation*}
  \mathbb{E}\left(\sqrt{A+B}\right) \approx \sqrt{\left(\mathbb{E}\sqrt{A} + \mathbb{E}\sqrt{B}\right)^2 - 2\mathbb{E}\sqrt{A}\mathbb{E}\sqrt{B}}.
\end{equation*}
Note that it is crucial to keep the expectation outside the square root of $B$.
With the mean-fluctuation decomposition in \eqref{mean_fluctuation_decomposition},
\begin{equation}\label{mean_fluctuation_expansion}
  u^2 + v^2 = (\overline{u} + u^\prime)^2 + (\overline{v} + v^\prime)^2 = (\overline{u}^2 + \overline{v}^2 + 2\overline{u}u^\prime + 2\overline{v}v^\prime) + (u^\prime)^2 + (v^\prime)^2.
\end{equation}
Denote by
\begin{equation}\label{Definition_AB}
  A := (\overline{u}^2 + \overline{v}^2 + 2\overline{u}u^\prime + 2\overline{v}v^\prime)\qquad\mbox{and}\qquad B := (u^\prime)^2 + (v^\prime)^2.
\end{equation}
Using the fact that the fluctuation has zero mean, it is clear that $\mathbb{E}\sqrt{A} = \sqrt{\overline{u}^2 + \overline{v}^2}$.

Next, both $(u^\prime)^2$ and $(v^\prime)^2$ satisfy chi square distribution with one degree of freedom
\begin{equation}\label{chi_square_u_v}
  (u^\prime)^2 \sim \sigma_u^2\chi_1^2\qquad\mbox{and}\qquad (v^\prime)^2 \sim \sigma_v^2\chi_1^2
\end{equation}
Since chi square distribution is the square of standard Gaussian distribution, there are prefactors $\sigma_u^2$ and $\sigma_v^2$ representing the variances of $u^\prime$ and $v^\prime$. 

For two random variables $\alpha$ and $\beta$ satisfying the chi square distributions with $m$ and $n$ degrees of freedom, namely, $\alpha\sim\chi^2_m$ and $\beta\sim\chi^2_n$,
\begin{equation}\label{Welch_Satterthwaite}
  \mathbb{E}\left(\sqrt{a\alpha+b\beta}\right)\approx \frac{\Gamma \left(\frac{(ma+nb)^2}{2(ma^2+nb^2) + \frac{1}{2}}\right)}{\Gamma\left(\frac{(ma+nb)^2}{2(ma^2+nb^2)}\right)}\sqrt{2\frac{ma^2+nb^2}{ma+nb}}
\end{equation}
which is the Welch-Satterthwaite equation \cite{van2019statistics}. In \eqref{Welch_Satterthwaite}, $\Gamma(\cdot)$ is the Gamma function. When $m=n=1$,
\begin{equation}\label{Welch_Satterthwaite_mn1}
  \mathbb{E}\left(\sqrt{a\alpha+b\beta}\right)\approx \frac{\Gamma \left(\frac{(a+b)^2}{2(a^2+b^2) + \frac{1}{2}}\right)}{\Gamma\left(\frac{(a+b)^2}{2(a^2+b^2)}\right)}\sqrt{2\frac{a^2+b^2}{a+b}}.
\end{equation}

\bibliographystyle{elsarticle-num}

\end{document}